\documentclass[preprint2]{aastex}
\usepackage{graphics}

\newcommand{\boomn}{{\sc Boomerang}}
\newcommand{\boom}{{\sc Boomerang} }

\begin{document}

\title{
Improved Measurement of
the Angular Power Spectrum of Temperature Anisotropy in the CMB
from Two New Analyses of \boom Observations
}

\author{
J.E.~Ruhl\altaffilmark{1},
P.A.R.~Ade\altaffilmark{2},
J.J.~Bock\altaffilmark{3},
J.R.~Bond\altaffilmark{4},
J.~Borrill\altaffilmark{5},
A.~Boscaleri\altaffilmark{6},
C.R.~Contaldi\altaffilmark{4},
B.P.~Crill\altaffilmark{7,8},
P.~de~Bernardis\altaffilmark{9},
G.~De~Troia\altaffilmark{9},
K.~Ganga\altaffilmark{10},
M.~Giacometti\altaffilmark{9},
E.~Hivon\altaffilmark{10},
V.V.~Hristov\altaffilmark{7},
A.~Iacoangeli\altaffilmark{9},
A.H.~Jaffe\altaffilmark{11},
W.C.~Jones\altaffilmark{7},
A.E.~Lange\altaffilmark{7},
S.~Masi\altaffilmark{9},
P.~Mason\altaffilmark{7},
P.D.~Mauskopf\altaffilmark{2},
A.~Melchiorri\altaffilmark{9},
T.~Montroy\altaffilmark{1,12},
C.B.~Netterfield\altaffilmark{13}, 
E.~Pascale\altaffilmark{6},
F.~Piacentini\altaffilmark{9},
D.~Pogosyan\altaffilmark{14,4},
G.~Polenta\altaffilmark{9},
S.~Prunet\altaffilmark{15,4},
G.~Romeo\altaffilmark{16}
}

\affil{
$^{1}$ Physics Department, Case Western Reserve University,
		Cleveland, OH, USA \\
$^{2}$ Dept. of Physics and Astronomy, Cardiff University, 
		Cardiff CF24 3YB, Wales, UK \\
$^{3}$ Jet Propulsion Laboratory, Pasadena, CA, USA \\
$^{4}$ Canadian Institute for Theoretical Astrophysics, 
		University of Toronto, Canada \\
$^{5}$ National Energy Research Scientific Computing Center, 
		LBNL, Berkeley, CA, USA \\
$^{6}$ IFAC-CNR, Firenze, Italy \\
$^{7}$ California Institute of Technology, Pasadena, CA, USA \\
$^{8}$ CSU Dominguez Hills, Carson, CA, USA \\
$^{9}$ Dipartimento di Fisica, Universita' La Sapienza, 
		Roma, Italy \\
$^{10}$ IPAC, California Institute of Technology, Pasadena, CA, USA \\
$^{11}$ Astrophysics Group, Imperial College, London, UK \\
$^{12}$ Dept. of Physics, UC Santa Barbara, CA, USA \\
$^{13}$ Depts. of Physics and Astronomy, University of Toronto, Canada \\
$^{14}$ Physics Dept., University of Alberta, Alberta, Canada \\
$^{15}$ Institut d'Astrophysique, Paris, France \\
$^{16}$ Istituto Nazionale di Geofisica, Roma,~Italy \\
}

\begin{abstract}
We report the most complete analysis to date of observations of
the Cosmic Microwave Background (CMB) obtained during the 1998 flight of
\boomn.  We use two quite different methods to determine the angular
power spectrum of the CMB in 20 bands centered at $\ell = 50$ to 1000, 
applying them to $\sim 50$\% more data than has previously been analyzed.  The
power spectra produced by the two methods are in good agreement with each
other, and constitute the most sensitive measurements to date over the
range $300 < \ell < 1000$.  The increased precision of the power spectrum
yields more precise determinations of several cosmological parameters
than previous analyses of \boom data.  The results continue to support
an inflationary paradigm for the origin of the universe, being well fit
by a $\sim 13.5$~Gyr old, flat universe composed of approximately 
5\% baryonic matter, 30\% cold dark matter, and 65\% dark energy, 
with a spectral index of initial density perturbations $n_s \sim 1$.
\\
\\
\end{abstract}

\section{Introduction}

Measurements of anisotropies in the cosmic microwave background (CMB)
radiation now tightly constrain the nature 
and composition of our universe.  High signal-to-noise
detections of primordial anisotropies have been made 
at angular scales ranging from the quadrupole~\citep{bennett96} to
as small as several arcminutes~\citep{mason02,pearson02,dawson02}.
The  power spectrum of temperature fluctuations shows a peak at
spherical harmonic multipole 
$\ell \sim 200$ which
has been detected with very high signal-to-noise by several 
teams~\citep{debernardis00,hanany00,halverson01,scott02}, and 
strong indications of peaks at higher $\ell$ have also
been found~\citep{halverson01,netterfield01,debernardis01}.  

Within the context of models with adiabatic initial 
perturbations, as are generally predicted by inflation,
these measurements have been used in combination with
various other cosmological constraints to estimate the values
of many important cosmological parameters.  Combining their 
CMB data with weak cosmological constraints such as a very loose 
prior on the Hubble constant, various teams have made 
robust determinations of several parameters, including the 
total energy density of the universe $\Omega_{total}$, 
the density of baryons $\Omega_b$, and the value of the 
density perturbation power spectral index, $n_s$
~\citep{lange01,balbi00,pryke01,netterfield01}.
Many other parameters are tightly constrained when stronger 
constraints on cosmology are assumed.

We report here new results from the 1998 Antarctic flight
of the \boom experiment.  Previous results from this flight
using less data than included here were published 
in \cite{debernardis00} (hereafter B00) and 
\cite{netterfield01} (hereafter B02).
Here we use the two very different analysis methods of B00 
and B02, and apply them  over a larger fraction of the dataset to 
make an improved measurement of the CMB angular power spectrum.  

\section{Instrument and Observations}

\boom is a balloon-borne instrument, designed to measure the 
anisotropies of the CMB at sub-degree angular scales.  The instrument
consists of a bolometric mm-wave receiver mounted at
the focus of an off-axis telescope, borne aloft on an altitude-azimuth
pointed balloon gondola.  Details of the instrument as it was
configured for the 1998 Antarctic flight, and its performance
during that flight, are given in \cite{crill02}.

The receiver consists of 16 bolometers, optically coupled 
to the telescope through a variety of cryogenic filters, feedhorns,
and reimaging optics.  We report here results from four of the six
150~GHz detectors in the focal plane, the same four analyzed
in B02.  The other two 150~GHz detectors exhibited non-stationary
noise properties and are not used in the analysis.  

The telescope has a 1.2~m diameter primary mirror and two cryogenic reimaging
mirrors mounted to the 2K surface of the receiver cryostat.  
These optics produce ($9.2'$,$9.7'$,$9.4'$,$9.5'$) FWHM beams at 
150~GHz in the four channels used here.  
The measured beams
are nearly symmetric Gaussians; the beamshapes are estimated
by a physical optics calculation, and calibrated by 
measurements on the ground prior to flight.
Uncertainty in the 
pointing solution ($2.5'$ rms) is estimated to smear the 
resolution of these physical beams to an effective resolution of
($10.9'$,$11.4'$,$11.1'$,$11.2'$) FWHM respectively.  
Based on the scatter of our various beam measures, and combined
with our uncertainty in the smearing due to the pointing solution
errors, we assign a 1-$\sigma$ uncertainty in the FWHM 
beamwidth of $1.4'$ in all channels.  This introduces an 
uncertainty in the measured amplitude of the power spectrum 
that grows exponentially with $\ell$ and that is correlated between
all bands.  This effect reaches a maximum of $\pm 40$\% 
in our highest bin ($\ell = 1000$), and is 
illustrated in Figure~2 of B02.   

The payload was launched from McMurdo Station, Antarctica
on 29 December 1998 and circumnavigated the continent 
in 10.5 days at an approximately constant latitude 
of -78~degrees.  During the flight, 247 hours
of data were taken, most of them on a ``CMB region" that 
was chosen for its very low dust contrast seen in 
the IRAS $100\mu$ maps of this region~\citep{moshir92}.

\begin{figure}[htb]
\begin{center}
\resizebox{3.0in}{!}{
  \rotatebox{90}{
  \includegraphics[0.5in,0.5in][8.0in,10.0in]{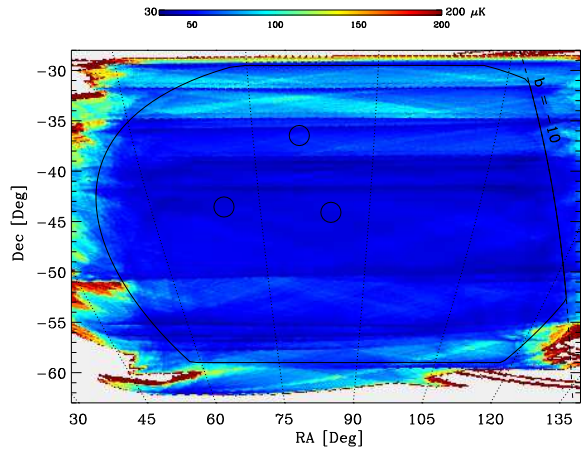}
  }
}
\end{center}
\caption{\small
The sky covered by CMB observations;  the color scale 
indicates the depth of coverage 
(diagonal component of the noise covariance matrix)
in a $7'$ Healpix pixel, in the map produced 
by the MADCAP analysis described below.
The region enclosed by the solid line is that used for the power
spectrum estimation.
The three circles show the locations of three bright
known quasars;  data within a $0.5^\circ$ radius of the quasars
is not used in the power spectrum estimation.  
\label{fig:mapnoise}
}
\end{figure}

The field observed in CMB scan mode 
is shown in Figure~\ref{fig:mapnoise}.  We analyze a subset
of this sky coverage here, chosen to be a contiguous region 
that is both sufficiently far from the galactic plane and 
well-covered by our observations.  Figure~\ref{fig:mapnoise}
shows the boundary of the region that we analyze in
this paper.  This region covers 2.94\% of the sky, and 
is defined as the intersection of:
\begin{itemize}
 \item an ellipse centered on RA = $88^\circ$, $\delta = -47^\circ$, with 
       semiaxes $a=25^\circ$ and $b=19^\circ$, where the
       short axis lies along the local celestial meridian,
\item the strip bounded by $ -59^\circ <  \delta  < -29.5^\circ $, 
\item and the region with galactic latitude $b < -10^\circ$.
\end{itemize}

This contour includes the best observed area of the survey, while
remaining far enough from the galactic disk to minimize 
galactic dust contamination.  It also does not have any small scale 
features (such as sharp corners) that could induce excessive 
ringing in the power spectrum extracted using one of our 
two methods (FASTER) discussed below.  This contour also excludes 
most of the scan turnarounds, where the scan speed is 
reduced and the low frequency noise can 
contaminate the angular scales of interest.

The vast majority of our observations of this region were 
made by fixing the elevation of the telescope and scanning 
azimuthally by $\pm 30^\circ$, typically centered roughly
$30^\circ$ from the anti-solar azimuth.  
Also used were the ``CMB region" 
portions of infrequently made ($\sim 1$ per hour) wider scans 
designed to traverse the Galactic plane as well.

CMB observations were made by scanning at three elevations 
($45^\circ$, $50^\circ$ and $55^\circ$), and at two 
azimuthal scan speeds (1 degree/second and 2 degrees/second,
hereafter 1dps and 2dps respectively).  The rising, setting
and rotation of the sky observed from $-78^\circ$ latitude
causes these fixed elevation scans to fill out the
coverage of a two dimensional map.  The color coding in 
our sky coverage map (Figure~\ref{fig:mapnoise}) gives the 
errors per pixel after coadding the
data from the four 150~GHz detectors.

The raw detector timestreams are cleaned, filtered and calibrated 
before being fed to the mapping and power spectrum estimation 
pipelines described below.  The cleaning and filtering used in this
analysis is identical to that described in B02 and is also
described in \cite{crill02};  we give the most relevant details here.

Bolometers are sensitive to any input that changes the detector
temperature, including cosmic ray interactions in the 
detector itself, radiofrequency interference (RFI), and thermal 
fluctuations of the baseplate heatsink temperature.
After deconvolving the raw bolometer data with the filter
response of the detector and associated electronics,
RFI, cosmic rays, and thermal events are found
by a variety of pattern-matching and map-based iterative
techniques.  Bad data are then flagged and replaced by
a constrained realization of the noise so that nearby data
can be used.  In the four channels used here, approximately 
4.8\% is flagged. 
The tails of thermal events are fit 
to an exponential and corrected,
and the data are used in the subsequent analysis.  Finally, 
a very low frequency high-pass filter is applied in the Fourier 
domain, with a transfer function 
$F(f) = 0.5 (1 - \cos(\pi f / 0.01 \mbox{Hz}) )$ for $f \le 0.01$~Hz,
$F(f) = 1$ for $f > 0.01$~Hz.

\section{Data Analysis Methods} \label{sec:DataAnalysis}

This paper reports our third analysis of data from the 1998 flight.
In B00 we reported the angular power spectrum found by analyzing 
data from a single detector covering 1.0\% of the sky, using roughly one 
detector-day of integration.
In B02 we reported results from four 150~GHz detectors, using
17 detector-days of integration on 1.9\% of the sky.
Here we report new results using 50\% more data from those same 
four detectors, using over 24 detector-days of integration on 
2.9\% of the celestial sphere.  

The results reported here use the same timestream cleaning and
pointing solutions described in B02.  In addition to the
larger sky cut, here we use
two independent and very different analysis methods 
which derive the angular power spectrum of the CMB from those 
timestream inputs.  One, using the MADCAP CMB analysis software
suite~\citep{borrill99},  creates a maximum likelihood map and
pixel-pixel covariance matrix from the input detector timestreams
and measured detector noise properties.  The power spectrum
is derived from the map and its covariance matrix;  this was
the method used in B00.  The other method, based on 
the MASTER/FASTER algorithms described in \cite{hivon02}
and \cite{contaldi02}, 
relies on a spherical harmonic transform of a filtered, simply binned
map created from those timestreams; the angular power spectrum in the
filtered map is related to the full-sky unfiltered angular power
spectrum through corrections derived from Monte-Carlos of the input
detector timestream and model CMB sky signals.  In the FASTER
procedure, the best fit angular power spectrum is then obtained by
using an iterative quadratic estimator analogous to that used in conventional
maximum likelihood procedures. This method was used in B02.

A theme of this paper is the comparison of the results from
these two very different analysis paths, and the stability of 
the cosmological results to any differences in the derived power spectra.

\subsection{Detector noise estimation}
Both MADCAP and FASTER require an accurate
estimate of the detector noise properties in order to determine
the angular power spectrum.  We estimate these noise
properties from the data themselves, using an iterative method 
to create an optimal, maximum likelihood map of the 
sky signal.  We then remove this signal from the
detector timestream prior to calculating the noise statistics.  This
method is described in both B02 and \cite{prunet01}.

For bolometer $i$ and iteration $j$, 
${\bf d}_i,{\bf A}_i$, ${\bf n}_i^{(j)}$, ${\bf N}_i^{(j)}$,
and ${\bf\Delta}^{(j)}$ are respectively the data,
pointing matrix, noise timestream, noise timestream correlation matrix,
and sky map.  The sky map and noise timestream correlation matrices
are found by iteration:

\begin{enumerate}

  \item Given the data timestream and estimated map, solve
	for the noise-only timestream with
	${\bf n}^{(j)}_i = {\bf d}_i - {\bf A}_i{\bf\Delta}^{(j)}$

  \item Use ${\bf n}^{(j)}_i$ to construct the noise timestream 
	correlation matrix, ${\bf N}_i^{(j)} = 
	\langle {\bf n}_i^{(j)}{\bf n}_i^{(j)\dagger} \rangle $ 

  \item Solve for a new version of the map using
	${\bf\Delta}^{(j+1)} =
	(\sum\limits_i {\bf A}_i^\dagger {\bf N}_i^{(j)-1} {\bf A}_i)^{-1}
	\sum\limits_i {\bf A}_i^\dagger {\bf N}_i^{(j)-1} {\bf d}_i$

  \item Return to step 1, using the new version of the map, and repeat.
	Iterate until the map ${\bf\Delta}$ and the 
	noise correlation matrices ${\bf N}_i$ are stable.

\end{enumerate}

For stationary noise ${\bf N}_i$
is diagonal in Fourier space, with the diagonal
elements equal to the power spectrum of the noise. 
We also assume, and check in practice, that the noise 
correlation between channels is negligible.

The noise correlation matrix ${\bf N}_i$ is
computed in Fourier space from the noise timestream
${\bf n}_i$ with a simple periodogram estimator. 
The maximum likelihood map 
of the combined bolometers, ${\bf\Delta}$,
is computed using a 
conjugate gradient approach~\citep{dore01},
which improves the recovery of large 
scale modes in the map.

Solving for all channels in a combined way takes advantage of the
redundant observations of the sky, therefore 
offering the best possible separation between signal and noise 
in the time streams for each bolometer.  The noise power spectrum 
estimation is well-converged after a few iterations,
typically three or four.

In this iterative procedure, we find a single 
maximum-likelihood map using all the data from 
all detectors.  In practice a separate noise covariance
is solved for in each of the 78 contiguous data ``chunks",
bordered by elevations moves or other timestream disturbances.
Thus, very slowly varying noise properties (eg a drift in the
instrument noise) will not affect the analysis.
Additionally, a line is evident in the noise power 
spectrum of the timestream data, varying slowly 
between 8 and 9 Hz over the course of the flight.  
We remove the effects of this non-stationary
source of noise by removing information in the 
timestream between 8 and 9 Hz.  These frequencies correspond 
to angular scales $\ell > 1000$, outside the range that 
we report here, for all scan speeds.

\subsection{The MADCAP Analysis Path}

Given a pixel-pointed time-ordered dataset $\bf d$ with piecewise
stationary Gaussian random noise, the maximum likelihood pixel map
${\bf \Delta}$ and pixel-pixel noise correlation matrix ${\bf C}_N$
are~\citep{wright96b,tegmark96b,ferreira00}
\begin{eqnarray}
  {\bf \Delta} & = & ( {\bf A}^\dagger {\bf N}^{-1} {\bf A} )^{-1}
		{\bf A}^\dagger{\bf N}^{-1} {\bf d} \nonumber \\
  {\bf C}_N &=&  ( {\bf A}^\dagger {\bf N}^{-1} {\bf A} )^{-1}
\label{madcapmapeqn}
\end{eqnarray}
where, as before, ${\bf A}$ is the pointing matrix and ${\bf N}$ is the
block-Toeplitz time-time noise correlation matrix.

Assuming that the CMB signal is Gaussian and azimuthally
symmetric, the maximum likelihood angular power spectrum $C_{l}$ is
that which 
maximizes the log-likelihood of the derived map given that 
spectrum~\citep{gorski94b, bond98},
\begin{equation}
{\cal L}( {\bf d} | C_{\ell}) = 
  - \frac{1}{2} \left( {\bf d}^\dagger \, {\bf C}^{-1} \, {\bf d} - {\rm Tr}
\left[ \ln {\bf C} \right] \right),
\end{equation}
where ${\bf C}$ is the full pixel-pixel covariance matrix. The
CMB signal and the detector noise are uncorrelated,
so ${\bf C}$ is just
the sum of the ${\bf C}_N$ found above, and the theory pixel-pixel 
covariance matrix ${\bf C}_T$ derived for a particular set of 
$C_\ell$'s.

In the MADCAP analysis path \citep{borrill99} we solve these equations 
exactly, calculating the closed form solution for the map,
using quasi Newton-Raphson iteration to find the set of $C_\ell$'s that
maximizes the log-likelihood~\citep{bond98}.  
Because the pixel-pixel correlation matrices are dense, the 
operation count scales as the cube, and the memory requirement as the
square, of the number of pixels in the map.  This imposes serious
practical constraints on the size of the problems we can tackle;
by optimizing our algorithms to minimize the scaling
prefactors, and using massively parallel computers, we have been able
to solve systems with up to O($10^5$) pixels - sufficient to analyze
this dataset at $7'$ pixelization.

There are two analyses that we want to perform on this dataset, 
each of which involve both map-making and power-spectrum
estimation.
First, we want to analyze the full dataset including all 
four channels at both scan
speeds, to solve for the CMB angular power spectrum.  Second, we 
want to perform a systematic test of the
self-consistency of the data,
differencing two halves of the data and checking that the sky signal
disappears.

For the first of these, we construct a time-ordered dataset by
concatenating the data from all four channels at both scan speeds,
and solve for the map using the eight associated 
time-time noise correlation functions.  
This timestream consists of 163,726,965 observations of
160,805 pixels.  
Using 400 processors on NERSC's 3000-processor IBM
SP3, the associated map and pixel-pixel noise correlation matrix can
be calculated from equation \ref{madcapmapeqn} in about 4 hours. Pixels not
included in the cut being analyzed here are removed from the map, and
the corresponding rows and columns of the pixel-pixel noise
correlation matrix are excised, equivalent to marginalizing over
them. The resulting map contains 92251 pixels, and the associated
noise correlation matrix fills 70 Gb of memory at 8-byte precision.

For the second analysis, we construct two time-ordered datasets, each
containing the data from all four channels but at only one of the two
scan speeds. 
The 2dps time-ordered data contains
74,879,196 observations over 124,257 pixels, while at 1dps we have
88,832,768 observations over 151,654 pixels. 
Having made the maps and pixel-pixel noise correlation matrices 
from each timestream we apply
our cut as above, and in addition remove any pixels within the cut
that are not observed in both halves of the flight. This results in
two maps (${\bf \Delta_A}$ and ${\bf \Delta_B}$) and their associated noise 
correlation matrices 
each covering an identical subset of 88,407 pixels.  
We then extract the power spectrum of the 
map ${\bf \Delta}_J = ({\bf \Delta}_A - {\bf \Delta}_B)/2$, 
taking the noise correlation matrix of ${\bf \Delta}_J$ to be 
the appropriately weighted sum of those for
the component maps, 
${\bf C}_N^J = ({\bf C}_N^A + {\bf C}_N^B)/4$.
This assumes the 2dps and 1dps noise are uncorrelated.  In
the power spectrum estimation process we 
assume a CMB-like theory correlation matrix when 
calculating the $C_\ell$'s.

The finite extent of our maps creates finite correlation
between our estimates of the power in nearby multipoles.  However, we
can reduce these correlations to small levels by calculating the
power in tophat bins of sufficient width.  We choose bins of width
$\Delta \ell = 50$, centered on $\ell = 50, 100, 150, ...1000$, together with additional ``junk" bins below $\ell = 25$ and above 
$\ell = 1025$,
which are included to prevent very low-$\ell$ and very 
high-$\ell$ power from being aliased into the range of 
interest.  This binning reduces the correlations between adjacent bins to
to less than $\sim 13$\% between neighboring bands.

The power in a multipole bin is related to the power in the individual
multipoles in that bin through a shape function: $C_{\ell} = C_{b}
C_{\ell}^{shape}$.  Although we are free to choose any spectral shape
within each bin, experience shows that for relatively narrow bins 
the particular choice makes very little difference.  We can 
explicitly account for the assumed
spectral shape in our cosmological parameter extraction; 
here we use a flat shape
function such that $\ell (\ell+1) C_{\ell}^{shape} = {\rm constant}$.

The maps derived from the time-ordered data have been smoothed by both
the detector beams and the common pixelization.  For constant,
circularly symmetric beams and pixels we can account for this exactly
by incorporating the appropriate multipole window function in the
pixel-pixel signal correlation matrix, ${\bf C}_T$.  

The fact that each
detector has a different beam means that ideally we should
construct individual maps and noise
correlation matrices for each channel and solve for the maximum
likelihood power spectrum of all four maps (each convolved with its
own beam) simultaneously.  However, this would give a 4-fold increase
in the number of pixels, and a 64-fold increase in the compute
time.  Instead, we analyze the single all-channel map
assuming a noise-weighted average beam; this approximation is 
quantitatively justified by tests done with the FASTER pipeline, below.

We use the HEALPIX pixelization~\citep{gorski98};  in 
this scheme, the pixels have 
equal area but are asymmetric and have varying shapes. 
These slightly different shapes lead to pixel-specific 
window functions. 
Calculating all of the individual pixel window functions at our
resolution is not feasible;
instead, we use the
all-sky average HEALPIX window function 
appropriate for our resolution.

By comparing individual pixel window functions at lower
resolution, we can set an upper limit on the
errors that may be induced by this approximation.  Scaling
to $27'$ pixels (a factor of 4 larger)
we find maximum deviations of 5\% in temperature (ie, 10\% in power) in the
ratio of actual pixel window functions to the average pixel window
function on the whole celestial sphere, at the
correspondingly scaled $\ell$ of 1024/4 = 256.
Thus, the pixel window function employed cannot be more
than 5\% (corresponding to an error in 
$C_\ell$ of 10\%) off the true value at our highest $\ell$.  
In fact, since the field incorporates pixels of many geometries, 
averaging will make the error much smaller, realistically less 
than 1\% in temperature.

Thus far we have assumed that the time-ordered data is comprised of
CMB signal and stationary Gaussian noise only.  However, we know that
in our data there are systematics that lead to residual
constant-declination stripes in the map.
Failure to account for these residuals leads 
to the detection of a signal in the (1dps-2dps)/2 
difference maps, which should be pure noise maps. 
In the MADCAP approach we account for these residuals by
marginalizing over the contaminated modes when 
deriving the power
spectrum from the map~\citep{borrill02}.  Specifically, 
to give zero weight to a particular pixel-template
we add infinite noise in that mode to the pixel-pixel
correlation matrix
\begin{equation}
{\bf C}^{-1}  \rightarrow  lim_{\sigma \rightarrow \infty} 
	\left( {\bf C} + \sigma^2 {\bf M}^\dagger {\bf M} \right)^{-1}
\end{equation}
where ${\bf M}$ is the matrix of orthogonal templates, one of 
each mode to be
marginalized over. Applying the Sherman-Morrison-Woodbury 
formula this reduces to
\begin{equation}
{\bf C}^{-1}  \rightarrow  {\bf C}^{-1} - {\bf C}^{-1} {\bf M} 
	\left( {\bf M}^\dagger {\bf C}^{-1} {\bf M} \right)^{-1} 
	{\bf M}^\dagger {\bf C}^{-1},
\end{equation}
yielding a readily calculable correction - requiring computationally
inexpensive matrix-vector operations only -- to the inverse
correlation matrix. 
Now whenever we multiply by ${\bf C}^{-1}$ in estimating
the power spectrum we simply add the appropriate correction term. For
the residual constant-declination stripes in this dataset we construct
a sine and cosine template along each line of pixels of constant
declination for all modes with wavelengths longer than 32 pixels
(about 4 degrees).

Once the iterative power spectrum estimation has converged,
the error bars on each bin are estimated from the initial (zero
signal) and final bin-bin Fisher information matrices using the offset
lognormal approximation \citep{bond98}.

\subsection{The FASTER Analysis Path}

The FASTER pipeline is based on the
MASTER technique described in
\cite{hivon02}.  MASTER allows fast and accurate determination
of $C_\ell$ without performing the time
consuming matrix-matrix manipulations that characterize exact methods
such as MADCAP~\citep{borrill99}.

As in MADCAP, there are two separate steps in the FASTER path;  
mapmaking, and 
power spectrum estimation from that map.  In our current implementation 
of FASTER, we make a map from the data by naively binning the timestream
into pixels on the sky.  To reduce the effects of $1/f$ noise on this
naively binned map, a brick-wall highpass Fourier filter is first 
applied to the timestream at a frequency of 0.1~Hz for the 1dps data,
and 0.2~Hz for the 2dps data.

The spherical harmonic transform of this naively
binned map is calculated using 
a fast ${\cal O}(N_{pix}^{1/2}\ell)$
method based on the Healpix tessellation of the sphere
\citep{gorski98}.  The angular power in a noisy map, 
$\tilde{C_{\ell}}$, can
be related to the true angular power spectrum on the full sky,
$C_{\ell}$, by the effect of finite sky
coverage ($M_{\ell \ell'}$), time and spatial filtering of the maps
($F_{\ell}$),
the finite beam size of the instrument ($B_{\ell}$), and instrument noise
(${N_{\ell}}$) as
\begin{equation}
\left\langle \widetilde{C_{\ell}}\right\rangle =
  \sum _{\ell'}M_{\ell\ell'}F_{\ell'}B_{\ell'}^{2}\left\langle
C_{\ell'}\right\rangle
+ \left\langle \widetilde{N_{\ell}}\right\rangle.
\label{eqn_cl}
\end{equation}
The coupling matrix $M_{\ell \ell'}$ is computed analytically.
$B_{\ell}$ is determined by the measured beam and the pixel window
function assuming here that the pixel has a circular symmetry.  $F_{\ell}$
is determined from Monte-Carlo simulations of signal-only time streams,
and $N_{\ell}$ from noise-only simulations of the time streams.

The simulated time streams are created using the actual flight pointing
and transient flagging.  The signal component of these time streams 
is generated from simulated CMB maps, while the noise component
is from realizations of the measured detector noise $n(f)$.  In both
cases the same high pass filtering 
(0.1~Hz at 1dps and 0.2~Hz at 2dps) and notch filtering 
(between 8 and 9~Hz, to eliminate the previously mentioned non-stationary
spectral line in the timestream data)
is applied to the simulated TOD as to the real one. $F_{\ell}$
and $N_{\ell}$ are determined by averaging over 600 and
750 realizations respectively.
Once all of these components are known the
power spectrum estimation is carried out as follows.

A suitable quadratic estimator of the {\em full sky}
spectrum in the {\em cut sky} variables ${\tilde C}_{\ell}$ together
with its Fisher matrix is constructed via the coupling 
matrix $M_{\ell \ell'}$ and the transfer function 
$F_{\ell}$ \citep{bond98,netterfield01}. The
underlying power is recovered through the iterative convergence of the
quadratic estimator onto the maximum likelihood value as in standard
maximum likelihood techniques. A great simplification and speed-up
is obtained due to the diagonality of all the quantities involved,
effectively avoiding the ${\cal O}(N^3)$ large matrix inversion problem
of the general maximum likelihood method.  
The extension of the quadratic estimator
formalism to Monte-Carlo techniques such as MASTER have the added
advantage that the Fisher matrix characterizing the uncertainty in the
estimator is recovered directly in the iterative solution and does not
rely on any potentially biased signal+noise simulation ensembles. A
detailed discussion of the FASTER extension to the MASTER procedure
can be found in \citet{contaldi02}.

A drawback of using naively binned maps in the pipeline is that the
aggressive time filtering completely suppresses the power in the maps
below a critical scale $\ell_c\approx 50$ \citep{hivon02}. This 
results in one or
more bands in the power spectrum running over modes with no power and
which are thus unconstrainable.  In practice we deal with this by binning
the power so that many of the degenerate modes lie within the first band
$2 < \ell \le 25$.  The power in the degenerate band can then be regularized
to zero power or a level consistent with the DMR large scale
results.  Regularizing with a non-zero value carries the disadvantage
that the second band $25 < \ell \le 75$ will be non-trivially correlated
with power which carries a theoretical bias.  Regularizing with zero
power results in no correlations between the first two bands and is
more consistent with the filtering done on the Monte-Carlo maps which
sets the signal in the affected modes to zero identically below
$\ell_c$.  We adopt the latter approach in this analysis to recover a
useful band which we label as $25 < \ell \le 75$.  However, as the
window functions will show below (Figure~\ref{fig:windowfunctions}), 
most of the information in this band comes from $50 < \ell \le 75$.

The FASTER pipeline allows the use of non-uniform masks 
applied to the observed patch of sky.
We have
experimented with a number of such weighting schemes for our patch
including total variance weighting $1/(S+N)$ and Weiner-like $S/(S+N)$
weighting where $S$ is the Monte-Carlo estimated variance of the 
coadded signal in the pixel (which varies from pixel to pixel 
because of the high pass filtering applied to the time data stream 
and the non-uniform scanning speed) and $N$
is the variance of the noise in the pixel.  We have found the $1/(S+N)$
weighting gives optimal results for this particular patch and coverage
scheme of this analysis.

In order to remove any effect of the constant-declination striping 
contaminant described
above a further (spatial) filtering step is
applied to all the maps in the pipeline.  The HEALPIX map is projected
to a rectangular, square-pixel map, where a spatial 
Fourier filter is applied that removes all modes in the map
with wavelengths greater than 8.2 degrees in the RA direction.
This filtered map is then projected back to the HEALPIX pixelization.

The inclusion of several channels is achieved by averaging the maps
(both from the data, and from the Monte-Carlos of each channel) before
power spectrum estimation.  Weighting in the addition is by hits per
pixel, and by receiver noise at 1Hz.  Each channel has a slightly
different beam size, which is taken into account in the
generation of the simulated maps. The Monte-Carlo procedure employed
in FASTER and MASTER ensures that the estimated power
is explicitly unbiased with respect to any known systematics, thus any
inaccuracy in assuming a common $B_{\ell}$ in the angular power
spectrum estimation is then absorbed into $F_{\ell}$. Similarly, any
inaccuracy on the effective pixel window function for the patch of sky
under consideration would be absorbed into $F_{\ell}$.

The calculation of the full angular power spectrum and covariance matrix
for the four good 150 GHz channels of \boom
($\sim 350,000$  $3.5'$  pixels and $\approx 216,000,000$ time samples;
this is different from the MADCAP numbers given above because 
non-CMB sections of the timestream are treated differently) 
takes approximately 
four hours running on six nodes of the NERSC IBM SP3.

\subsection{Application to Data}

When treating real data, each of the methods described 
has particular advantages.  MADCAP is
an ``optimal" tool, in the sense that it uses the full
statistical power of the data in deriving the power spectrum; 
no other method can use the same data and
produce a power spectrum with smaller errorbars.  
In addition, it produces a maximum-likelihood map and
pixel-pixel covariance matrix that take advantage of the
full cross-linking of the scan strategy.  However,
the MADCAP algorithms are computationally very costly.
This leads to the use of several approximations ({\em eg} the 
use of a single beam for the four channels, using an 
average pixel window function, and fitting errors as a 
lognormal function), 
and reduces our ability to use this method for wide-ranging
testing of potential systematic effects.
With our current computing power and sky cut, we are limited to 
a $7'$ pixelization with MADCAP.  

The FASTER method provides a less-optimal
estimate of the power spectrum, but is computationally
much more rapid.  As will be shown below, in our case
the FASTER results are nearly as statistically 
powerful as those from MADCAP.  The rapid computational turnaround
allows the use of finer pixelization ($3.5'$), and
extensive systematic testing and modeling of potential systematic
errors.  Additionally, FASTER is capable of handling 
independent beams, and enables the computation of a 
true window function for our $\ell$ bins for use in 
parameter extraction.

\section{Signal maps}

\begin{figure}[htb]
\begin{center}
\resizebox{3.0in}{!}{
  \rotatebox{90}{
  \includegraphics[0.5in,0.5in][8.0in,10.0in]{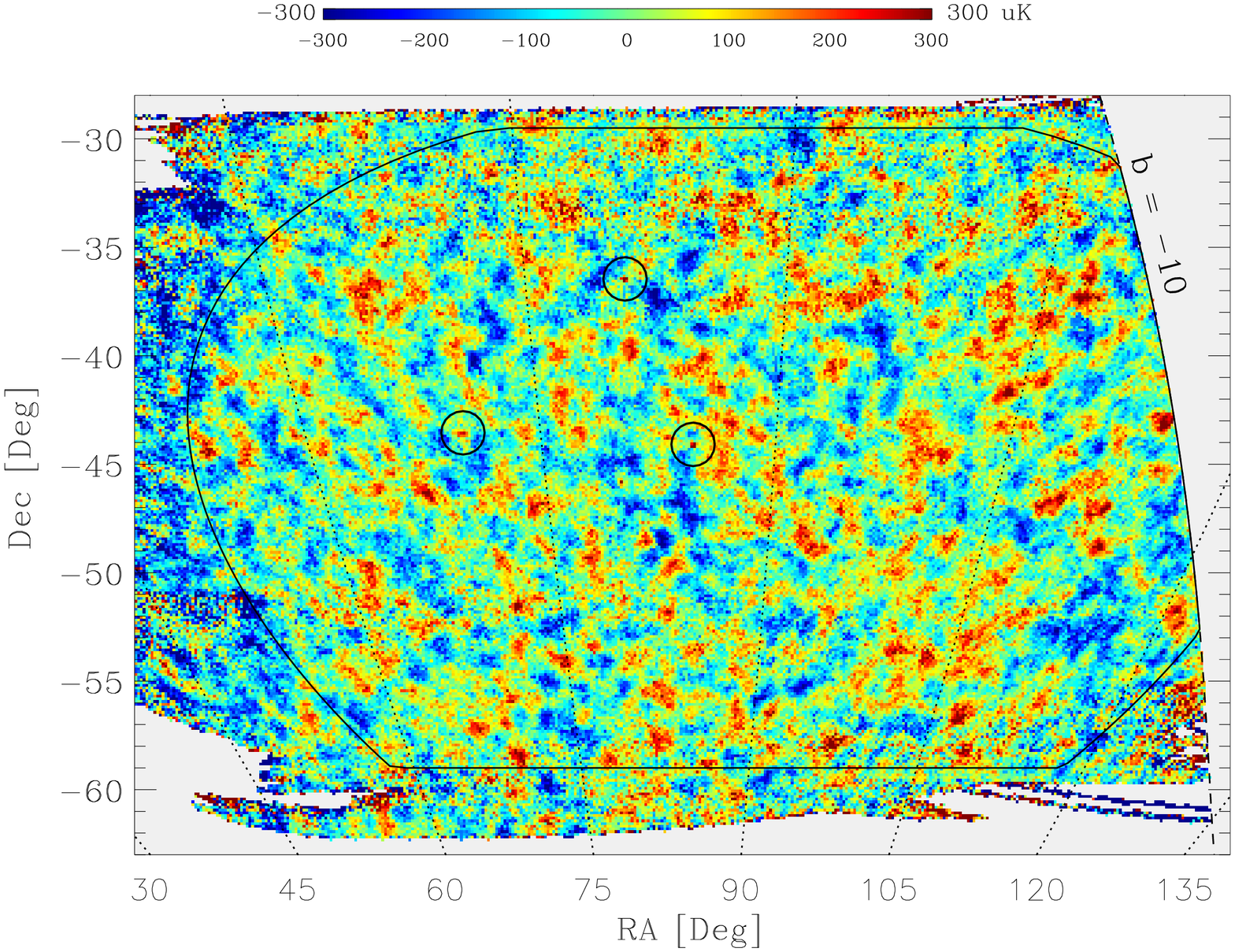}
  }
}
\resizebox{3.0in}{!}{
  \rotatebox{90}{
  \includegraphics[0.5in,0.5in][8.0in,10.0in]{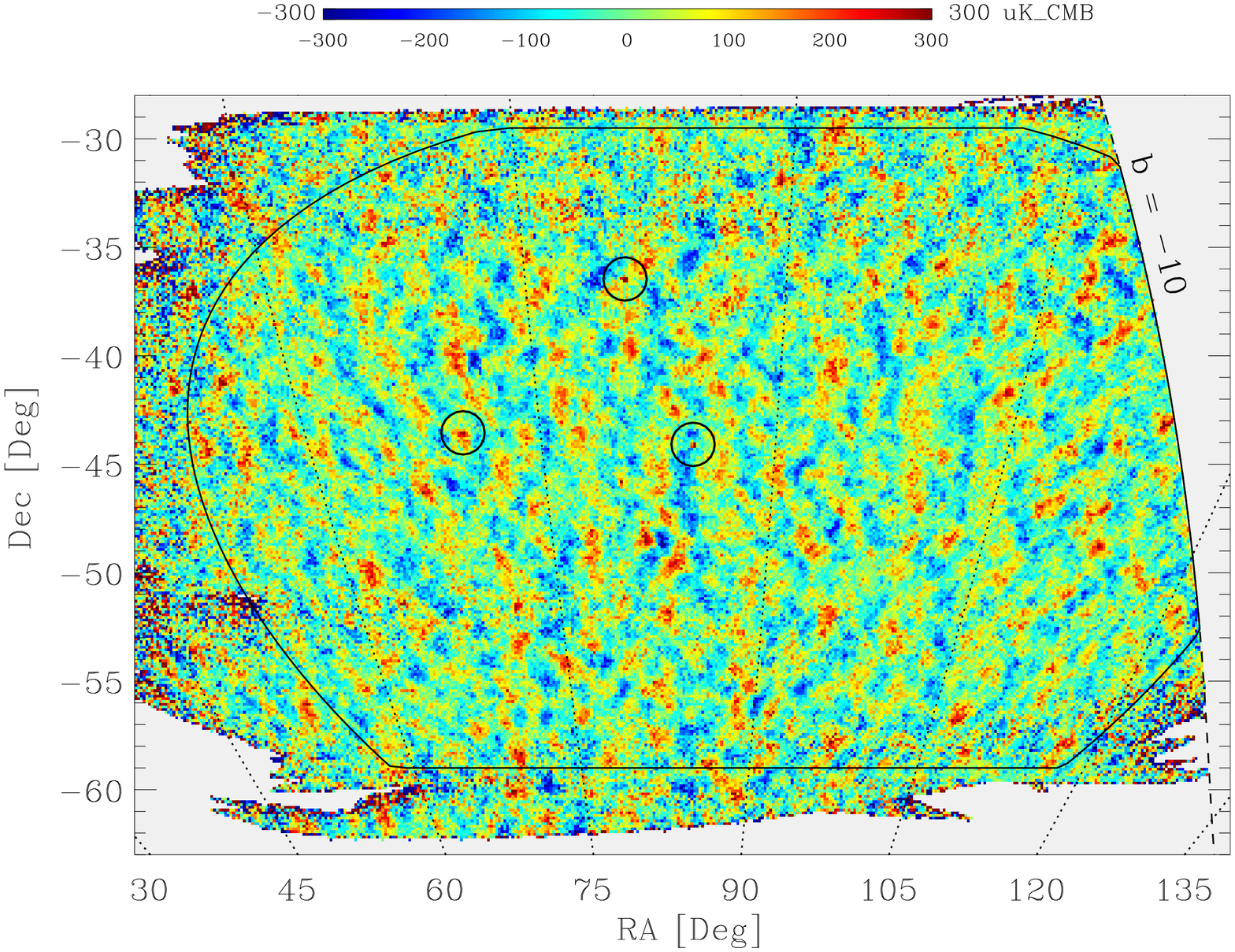}
  }
}
\end{center}
\caption{\small
The maps of CMB temperature produced by MADCAP (top) and
FASTER (bottom).  For comparison, both maps are 
pixelized at $7'$; in practice we use a $7'$ ($3.5'$) pixelization in the 
MADCAP (FASTER) analysis.
The strikingly different appearance of the
maps, with the MADCAP map preserving more information on large scales, 
illustrates some of the significant differences in the two
analysis methods, as described in the text.
\label{fig:cmbmaps}
}
\end{figure}

The first step in each pipeline is the production of a sky map.  
The fundamental differences between the two analysis paths
are well illustrated by a visual comparison of the two maps, shown in
Figure~\ref{fig:cmbmaps}. 
Though there is a high correlation of the small-scale
structure in these two maps, their overall appearance is
strikingly different, due primarily to the time-domain filtering that
suppresses large scale structure in the FASTER map.  In addition, the
FASTER map has had the constant declination modes removed 
(hereafter, ``destriped"), while the MADCAP map has not.  (The
MADCAP destriping occurs via marginalization over contaminated modes
during the power spectrum estimation).  The MADCAP map should 
be interpreted in concert with its covariance matrix, which
describes which modes in the map are well constrained and which are
not.  The FASTER procedure does not create a covariance matrix; the
correlations in the map are accounted for in the
Monte-Carlo process.  Thus while it is reassuring that the two maps
show similar structure on small scales, a quantitative comparison can
only be made by proceeding through the estimation of the angular
power spectrum with each method.

\section{FASTER Analysis consistency tests}

\begin{figure}[htb]
\begin{center}
\resizebox{3.0in}{!}{
  \includegraphics[0.5in,2in][8.0in,11in]{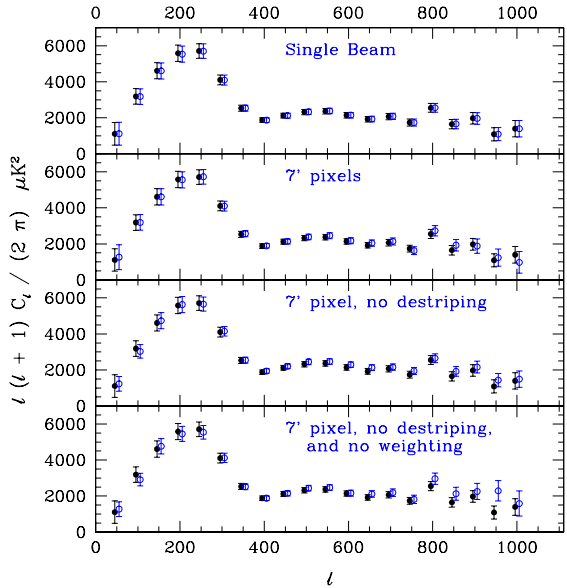}
  }
\end{center}
\caption{\small
Angular power spectra derived from the FASTER pipeline.   
The solid black circles in each panel show the reference
FASTER spectrum, which is derived from a $3.5'$ pixelized
map using S+N weighting, spatially filtered as described in the text to 
remove constant-declination stripes.
In the top panel the reference spectrum is compared with a 
spectrum derived using a single beam window function, as in MADCAP.
The second panel shows the effect of using $7'$ pixelization.
The third panel illustrates the effect of removing the constant
declination stripes;  the primary effect is to increase the error
in the first bin.  The bottom panel shows the result of using
a uniformly weighted map and neglecting to remove the constant 
declination stripes.  The top three panels show excellent agreement 
with the reference spectrum, while the bottom panel shows good
agreement except at very high $\ell$.  
\label{fig:faster4panel}
}
\end{figure}

The MADCAP analysis is limited to $7'$ pixelization, and assumes
a common beam window function for the four channels.  We have
used the FASTER pipeline to check the effect of this coarser
pixelization and window function assumption with respect to the
baseline FASTER result, which is calculated using $3.5'$ pixels and
individual window functions for each channel.  The baseline FASTER
result is shown in Figure~\ref{fig:faster4panel},
which also gives results derived using $7'$ pixels, and results
derived using the same ``single beam" assumption used by MADCAP.
As can be seen in the figure,
the single beam approximation has negligible effect.  The 
$7'$ pixelization does have some effect, but it is small
compared with the statistical errors.

We have also tested the robustness of the FASTER result
to other changes in the pipeline.
Figure~\ref{fig:faster4panel} also shows the effects of destriping, and
of using signal+noise weighting (rather than uniform weighting).
These have some affect on the power spectrum, 
again smaller than the statistical errors.  Note that we expect
these to have some effect given that the information content of the
map is modified by these procedures. 

Another test of the robustness of the angular power spectrum
is to change the details of the $\ell$ binning.  We have used the FASTER
pipeline to derive power spectra with 
$\Delta \ell = 40$ bins, and for $\Delta \ell = 50$ bins shifted
by 25 from our fiducial binning.  Both of these give
excellent agreement with power spectrum of Figure~\ref{fig:faster4panel}.  

\section{Internal Consistency Tests}

The analysis pipelines described above deliver an estimated
CMB power spectrum along with statistical errors on that power
spectrum.  Below, we will show the CMB power spectra derived
from the maps, and use those results to estimate cosmological 
parameters.  Before doing so, we describe here a variety of 
internal consistency tests designed to check for residual 
systematic contamination.

Our internal consistency checks are done by splitting the 
dataset roughly in half, making a map with each half of the data,
subtracting these two maps, and asking whether the power
spectrum of the residual map is consistent with pure detector noise.
Note that one only {\sl expects} the two maps to be identical
if they contain the same information;  if the two maps 
have been observed or filtered differently, perfect agreement is not 
expected.  

Our most powerful internal consistency check is to take data 
that was gathered while scanning the gondola azimuthally at
1dps (roughly the first half of the flight)
and compare it with data taken during 2dps scans
(roughly the second half of the flight).  This tests for
effects that vary over long timescales, position of the gondola
over the Earth, position of the scan region with respect to the Sun,
and instrumental effects that are modulated by scan speed.
The latter include any mis-estimate of the transfer function
of the detector system and any non-stationary noise 
in the detector system.  Hereafter, this test is
referred to as the (1dps-2dps)/2 consistency test.  Each pipeline was used
to produce and estimate the power spectrum of a (1dps-2dps)/2 map.

\begin{figure}[htb]
\begin{center}
\resizebox{3in}{!}{
  \rotatebox{90}{
  \includegraphics[0.5in,0.5in][8.0in,10.0in]{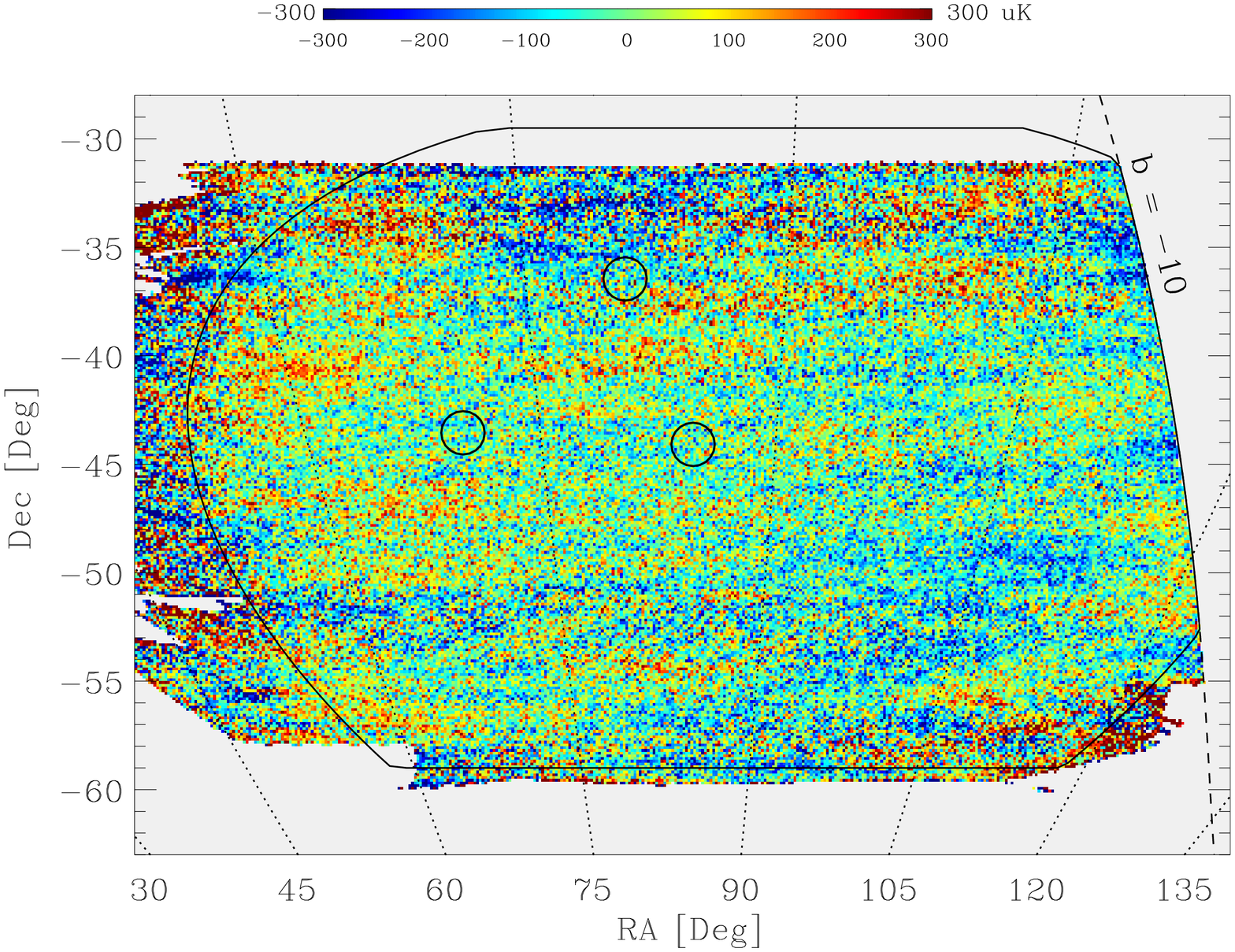}
  }
}
\resizebox{3in}{!}{
  \rotatebox{90}{
  \includegraphics[0.5in,0.5in][8.0in,10.0in]{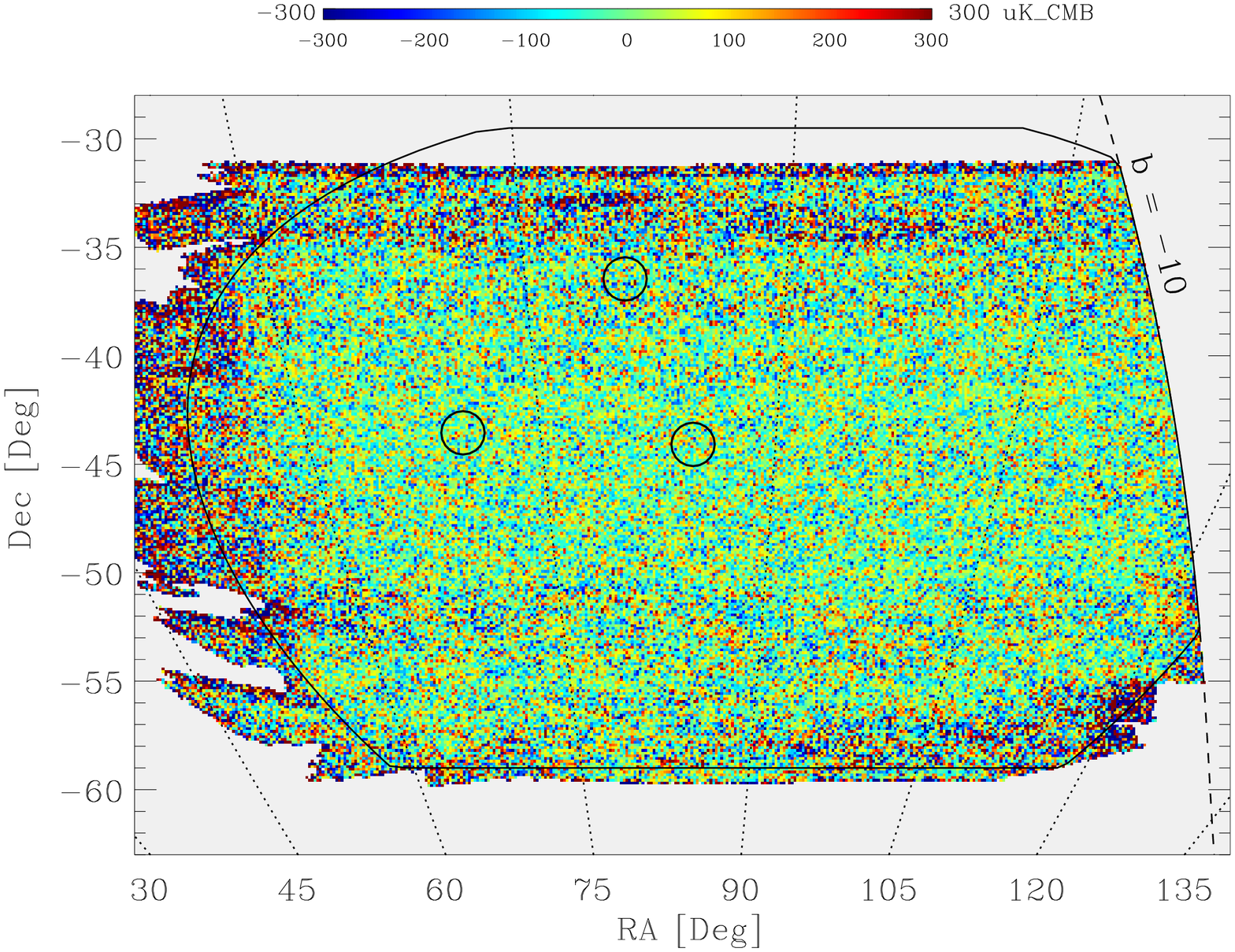}
  }
}
\end{center}
\caption{\small
(1dps-2dps) difference maps, both at $7'$ pixelization to facilitate
map comparisons by eye, for the region of sky where these scans
overlap.  The color scale is the same as for the previous figures.
Note that the consistency test power spectra are calculated on these maps
divided by two, (1dps-2dps)/2.
Top panel:  The MADCAP difference map.  This map is not
destriped, since in that pipeline the constant-declination stripes are 
ignored (by introducing a constraint matrix) in the derivation of 
the angular power spectrum.
Bottom panel: The destriped FASTER difference map.  Note that the 
MADCAP input timestream contains additional low frequency information
that is removed by an additional highpass filter in the 
FASTER pipeline.
\label{fig:jackmaps}
}
\end{figure}

Figure~\ref{fig:jackmaps} shows 
MADCAP and FASTER (1dps-2dps) difference maps, each
pixelized at $7'$. 
Many of the gross features apparent in both maps are due to
the variations in S/N, as can be seen by comparison with 
Figure~\ref{fig:mapnoise}.
The MADCAP map is not destriped, because
the destriping in that pipeline is done with a constraint matrix in deriving
the power spectrum.  The FASTER map is destriped; and appears
significantly cleaner to the eye.  In practice, a $3.5'$ pixelized map 
is used in the FASTER analysis;  here we display a $7'$ map so the noise
level per pixel remains comparable to the MADCAP version.

\begin{figure}[htb]
\begin{center}
\resizebox{3in}{!}{
  \includegraphics[0.5in,2.0in][8in,10in]{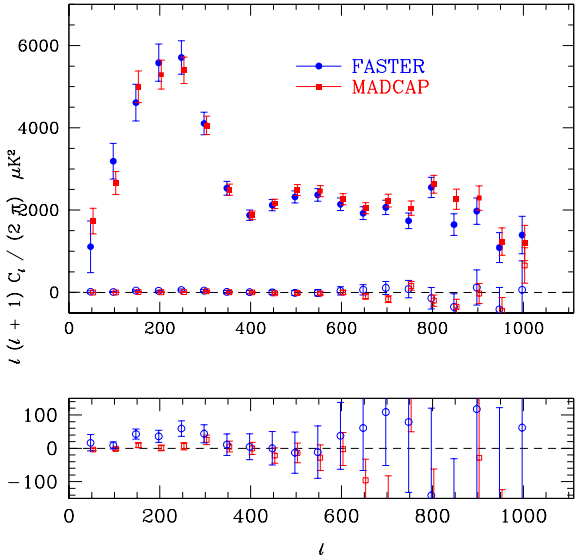}
}
\end{center}
\caption{\small
MADCAP and FASTER angular power spectra and (1dps-2dps)/2 difference map
power spectra.  Top Panel:  the FASTER (filled blue circles) and MADCAP 
(filled red squares) angular power spectra, and their
respective (1dps-2dps)/2 difference map power spectra (open symbols).
The effects of constant-declination stripes have been removed 
in each of these analyses.  Bottom panel: the difference map angular 
power spectra plotted on a magnified scale.  There is 
a systematic effect near $\ell \sim 200$ in the FASTER power 
spectrum, which is absent in the MADCAP treatment.  
The level of these residuals is much smaller than the 
statistical errors on the full power spectrum, shown in the top panel.
\label{fig:jackPS}
}
\end{figure}

Figure~\ref{fig:jackPS} shows the power spectra of the signal maps 
(top panel) shown in Figure~\ref{fig:cmbmaps} and 
of the (1dps-2dps) difference maps 
(bottom panel) shown in Figure~\ref{fig:jackmaps}.  It 
is apparent that the power spectra of the
signal maps are in very good agreement with one another; these are
discussed in more detail below.  Here we focus on the (1dps-2dps)/2
difference spectra.

The statistical error in the power spectra of the signal maps
is dominated by sample variance for $\ell < 500$.  Because there is no
signal and thus no sample variance in the power spectra of 
the difference maps, the
difference maps are sensitive to systematic effects that are well
below the (sample variance dominated) statistical noise of the signal
maps at low $\ell$.

The FASTER Monte-Carlo simulations show that the
different scanning and $\ell$-space filtering in the 1dps 
and 2dps data leads to a leakage of CMB signal into 
the (1dps-2dps)/2 FASTER map.  The average level of this signal is 
expected to be at the level of $\sim 10 \mu \mbox{K}^2$ 
near the first peak at $\ell \sim 200$.  We correct for this
effect in the FASTER pipeline consistency test by subtracting
the Monte-Carlo mean residual power found in each bin 
from the actual (1dps-2dps)/2 power spectrum,
and by adding the variance of this effect 
in quadrature to the errors
on that power spectrum.  

After these corrections to the FASTER pipeline, we find 
the difference map angular power spectra shown in 
the bottom panel of Figure~\ref{fig:jackPS}.
The $\chi^2$ per degree of freedom
with respect to a zero-signal model is 1.34 (1.28) with
a probability of exceeding 
this $\chi^2$ of $P_> = 0.14$ (0.18) for the MADCAP (FASTER)
analysis, respectively.  
Thus, when the entire spectrum is
considered, the difference spectra of both analysis methods are
reasonably consistent with zero.  It is clearly apparent, however,
that there is a statistically significant signal in the FASTER
difference spectrum, at $\ell \le 300$. 
Over this limited range of the
spectrum, the FASTER spectrum has a reduced
$\chi^2 = 3.7$ for 6 degrees of freedom, for a $P_> = 0.001$.
Over the same range, the MADCAP
analysis gives a reduced $\chi^2 = 1.10$ for 6
degrees of freedom, for a $P_> = 0.36$.

The residual signal in the FASTER difference map is both
localized in $\ell$ 
and very small, with a mean of only $45 \mu\mbox{K}^2$ in the four
bins $150 < \ell < 300$.  The CMB
signal is roughly 5000 $\mu\mbox{K}^2$ in this $\ell$ range,
and our statistical errors on the CMB signal, dominated by sample
variance, are $\sim 400 \mu\mbox{K}^2$.  Thus, though the 
FASTER pipeline formally fails this 
test, our statistical errors dominate our systematic 
errors by an order of magnitude.

Investigation of individual detector channels shows that 
the (1dps-2dps)/2 power spectra near $\ell \sim 200$
are of similar shape and amplitude in each.
We have done a variety of other consistency tests and simulations 
using the FASTER pipeline on our lowest-noise channel, B150A,
to try and understand potential sources for the (1dps-2dps)/2 
failure.  We have broken the data into four quarters (Q1 and Q2 at 1dps;
Q3 and Q4 at 2dps) and found difference map power spectra for
combinations that minimize effects that depend on scan speed
[(Q1+Q3)-(Q2+Q4)] or a drift in time [(Q1+Q4)-(Q2+Q3)].  These 
combinations fail the consistency test with amplitudes and shapes similar 
to the (1dps-2dps)/2 failure.

Simulations were done in an attempt to recreate the 
(1dps-2dps)/2 difference failure by inducing various systematic effects.
Changes in the gain, the pointing offset, and the filtering were
modeled.  Of these, only the last can explain the failure in the
FASTER pipeline, given that the data pass the test in the MADCAP
pipeline, since gain and pointing offsets should be treated identically
by the two methods.  For plausible levels of these systematic errors,
none induced (1dps-2dps)/2 failures at the level seen.  The
systematic that created the most similar shape was a pointing offset
between the two data sets.  This is not 
a priori unlikely, as it is plausible that a differential 
offset might occur in the attitude reconstruction for the two 
scan speeds.  The magnitude of the difference test failure would correspond 
to an $\sim 7'$ offset between the two
data sets.  This is inconsistent with the measured stability of the
positions of the quasars in the two maps and, more importantly, is
inconsistent with the fact that the MADCAP analysis achieves equally
high or higher sensitivity and passes this test.  We have
not been able to find the cause of the FASTER analysis failure of 
the (1dps-2dps)/2 consistency test.

\begin{figure}[htb]
\begin{center}
\resizebox{3in}{!}{
  \includegraphics[0.5in,2.0in][8in,10in]{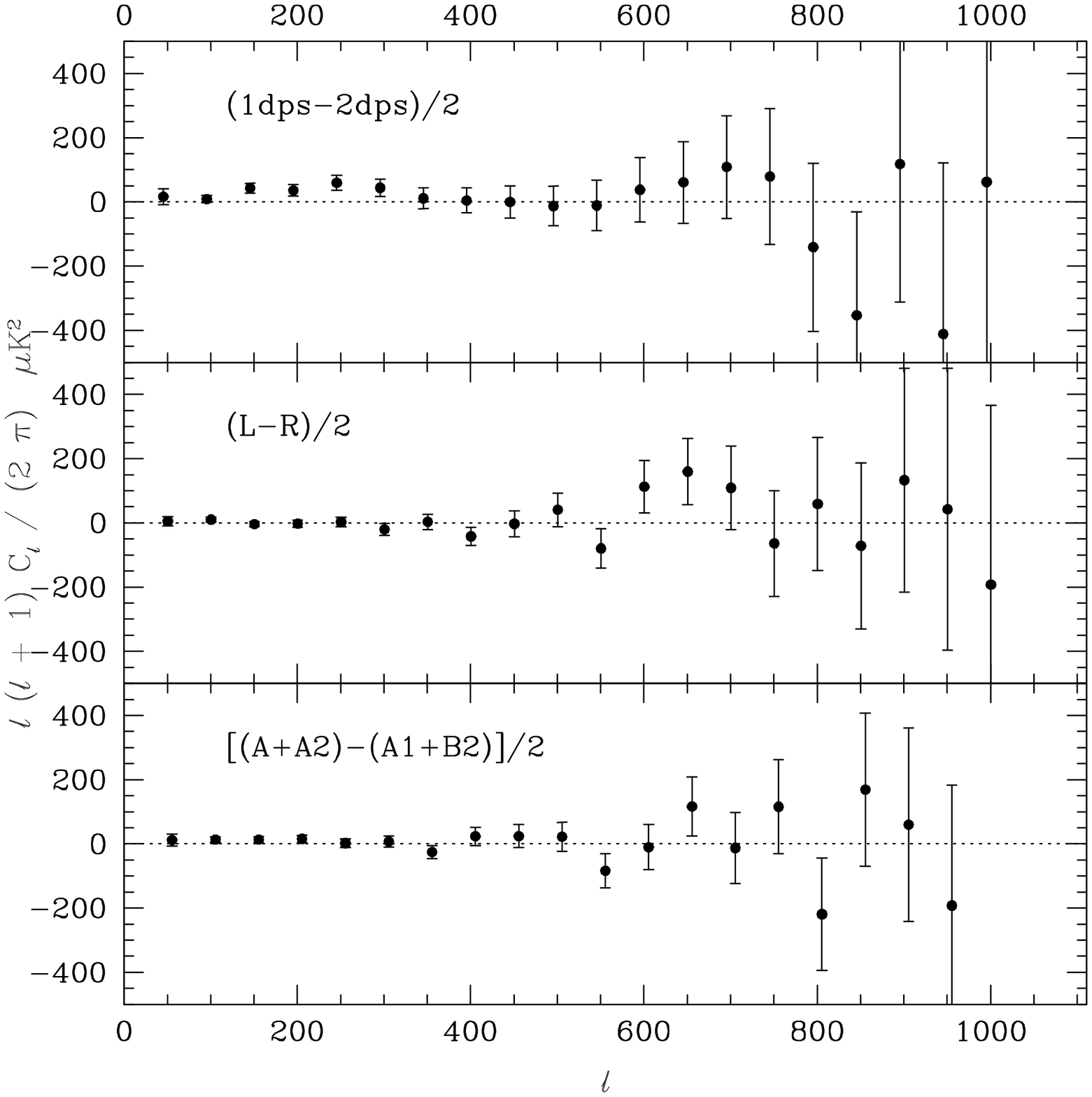}
}
\end{center}
\caption{\small
Difference map power spectra derived using FASTER.  
The top panel shows the (1dps-2dps)/2 difference map results.
The middle panel shows the result found by difference maps
made from leftgoing scans and rightgoing scans, (L-R)/2.
The bottom panel is for a map made by 
differencing maps made by two channel combinations, [(A+A2)-(A1+B2)]/2.  
While the latter two power spectra are relatively consistent with zero
contamination, near $\ell \sim 200$ the (1dps-2dps)/2 spectrum is not.
This contamination is, however, much smaller than the statistical errors 
on the full CMB power spectrum in this $\ell$-region.
The $\chi^2$ statistics of these spectra with respect to zero signal 
are given in Table~\ref{tab:jacks}.
\label{fig:3jacks}
}
\end{figure}

We also used the FASTER pipeline to perform two other consistency 
tests on the real data.  These are shown, along with the 
(1dps-2dps)/2 results
for reference, in Figure~\ref{fig:3jacks}.  One
differences maps made using rightgoing vs. leftgoing scans.
Another compares maps made with two of the four channels
(channels A and A2) with the other two (channels A1 and B2).
Both of these power spectra appear to be consistent with zero
in all $\ell$ regions, as evidenced by the statistics quoted in 
Table~\ref{tab:jacks}.

\begin{deluxetable}{lccc}
\tabletypesize{\scriptsize}
\tablecaption{Consistency Test Results \label{tab:jacks}}
\tablewidth{0pt}
\tablehead{
\colhead{Test} & 
\colhead{bins} & 
\colhead{Reduced $\chi^2$}   & 
\colhead{P$_>$}
}
\startdata
FASTER (L-R)/2 & all & 1.15 & 0.29 \\
	       & 1-6 & 0.96 & 0.45 \\ \\
FASTER [(A+A2)-(A1+B2)]/2 	& all & 1.18 & 0.26 \\
	     			& 1-6 & 1.25 & 0.28 \\ \\
FASTER (1dps-2dps)/2 		& all & 1.28 & 0.18 \\
	     			& 1-6 & 3.70 & 0.001 \\ \\
MADCAP (1dps-2dps)/2 		& all & 1.34 & 0.14 \\
	     			& 1-6 & 1.11 & 0.35 \\
\enddata

\tablecomments{\small
$\chi^2$ statistics for the consistency tests described in 
the text and plotted in Figures~\ref{fig:3jacks} 
and \ref{fig:jackPS}.  Results are reported for all $\ell$ bins 
($50 \le \ell \le 1000$, 20 bins) as well as for the 
first six bins 
($50 \le \ell \le 350$), where the FASTER (1-2dps)/2 test failure of 
Figure~\ref{fig:3jacks} is evident.
}
\end{deluxetable}

The (1dps-2dps)/2 test
failure on the FASTER pipeline leads us
to the inclusion of an additional systematic error term in 
the region
where that failure is significant, ie for $\ell \le 400$.  
In our final results below, we increase the quoted FASTER errors 
on those bins by the amount of the failure, adding 
it in quadrature (in $\mu\mbox{K}^2$) to the likelihood derived errors.
In the Fisher matrix this corresponds to adding the difference map
power spectrum residuals in 
quadrature to the diagonal elements, while leaving the off-diagonal terms 
unmodified.

\section{Comparison of results}

The discussion above leads us to believe that the larger pixels
and the single-beam approximation used by MADCAP should not
have a significant effect on the power spectrum.  In addition,
we have learned of a small consistency test failure over a small 
range of $\ell$ in the FASTER power spectrum, and 
corrected the errors on the spectrum accordingly.

We now turn to the comparison of the CMB power spectra derived
with FASTER and MADCAP, shown in 
the top panel of Figure~\ref{fig:jackPS}.

Despite the fact that they were derived from the same timestream
data, there are several reasons why these two power spectra 
are not expected to be identical.  
Both the crosslinked observing strategy and the lower frequency
filtering cutoff in the timestream allows MADCAP to 
recover some modes that are missing from the FASTER map.  

At the level of the errors shown, the agreement between these 
two estimates of the power spectrum is excellent.  
However, there is some indication of a 
systematic ``tilt" between the two spectra.  The level of this tilt is 
not large;  modeling it as a difference in the beam window functions, 
reducing the FWHM of the beam used by MADCAP by one quarter of 
our systematic beam uncertainty, visually removes the apparent tilt.
For this reason, and as is borne out by the discussion below, 
this difference will not have much effect on the cosmological 
parameter estimation results.

However, we have investigated any known differences that could
lead to a systematic difference between these two power estimation methods.
We have shown (via the FASTER consistency tests discussed above) 
that the larger pixelization and single-beam assumption of MADCAP should
not produce such a tilt.  Another potential effect is a bias
in the pixel window function, which MADCAP takes to be the 
average HEALPIX window function on the sphere.  The
FASTER Monte-Carlos incorporate
the effects of the real pixel geometries;  any bias induced by
using a single, isotropized approximation for the smoothing of the
HEALPIX pixelization is corrected by Monte-Carlo estimation of the
transfer function $F_{\ell}$.  In effect the transfer function ensures
the method is robust to any similar approximations used in describing
the effective pixelization smoothing.  However, the analytic arguments
discussed above, based on individual pixel window functions calculated
for larger pixels, indicate that any such bias caused by the MADCAP 
assumption should be very small.

Another potential bias could be introduced by the 
destriping algorithms.  We have used Monte-Carlos to test 
for such effects in FASTER, and have found that any such bias is
much smaller than the effect seen here.  The marginalization 
method used by MADCAP is not expected to bias the power spectrum
in any way, but Monte-Carlo tests to verify this are not 
practical given the greater computational cost of that
pipeline. 

In principle a tilt could also be induced by a difference 
in the timestream noise
statistics used by one of the methods;  however, the same noise 
power spectrum (or time-time noise correlation function) is used by the
two pipelines.

It is possible that the constant declination striping is not fully 
removed by one of the destriping algorithms, and this leads 
to the difference in tilt.  As can be seen in 
Figure~\ref{fig:faster4panel}, the FASTER destriping does affect the
power spectrum slightly at high $\ell$.  If this is
the reason for the tilt discrepancy, residual striping that 
is randomly phased with respect to the CMB sky signal would increase
the level of the power spectrum.  

\begin{figure}[htb]
\begin{center}
\resizebox{3.1in}{!}{
  \includegraphics[0.5in,2.0in][8in,10in]{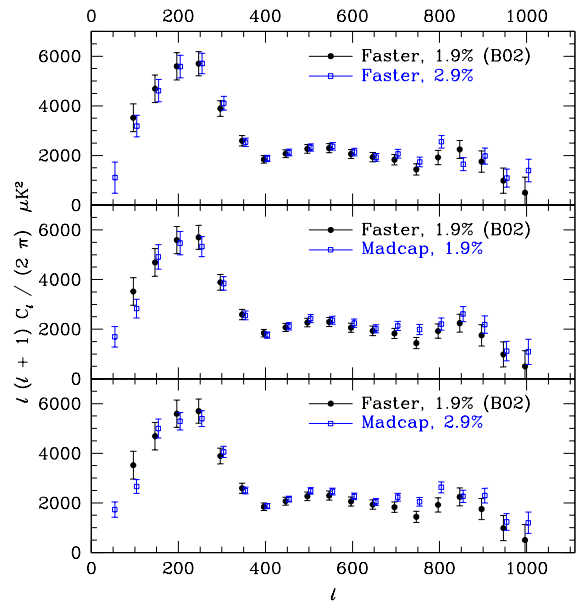}
}
\end{center}
\caption{\small
A comparison of the FASTER results of B02 (black filled circles)
derived from 1.9\% of the sky at $7'$ pixelization, with three new 
analyses (open blue squares in each panel).  
Top panel: the FASTER results of this paper (2.9\% of the sky, 
$3.5'$ pixelization).  
Middle panel: a new MADCAP analysis of the B02 sky cut 
(1.9\% of the sky, $7'$ pixelization).  
Bottom panel: the MADCAP results of this paper (2.9\% of the sky,
$7'$ pixelization).
The agreement is generally very good, with the greatest variations at
high $\ell$ where noise, rather than cosmic variance, dominates the 
errors.
\label{fig:nettcompare}
}
\end{figure}

Figure~\ref{fig:nettcompare} compares the B02 result, derived using
FASTER on 1.9\% of the sky at $7'$ pixelization, with 
several new results.
The top panel compares the B02 result with the 
final FASTER result discussed above, on 2.9\% of the sky
at $3.5'$ resolution.  The middle panel shows a new MADCAP 
analysis of the same region as B02, with the same resolution.
Finally, in the bottom panel B02 is compared with the MADCAP 
analysis of the larger cut analyzed in this paper, at $7'$ resolution.
As expected, the larger dataset
leads to smaller errorbars across the entire range of $\ell$.  
The MADCAP errors are smaller than the FASTER errors at low
$\ell$, due to the preservation of lower frequencies in the timestream.
The FASTER results agree very well with one another, except 
in the region near $\ell \sim 800$ where there are three 1-$\sigma$
and one 2-$\sigma$ deviations.  In the lower panels there is some 
evidence for the same tilt bias between MADCAP and FASTER on 
the B02 cut (mentioned above), indicating this is not unique to the 
larger sky cut.

\section{Galactic Dust}

In \cite{masi01} we 
measured the angular power spectrum of the \boom 
410~GHz map in three circles of 
$9^\circ$ radius, centered at galactic latitudes of
$b = -38^\circ$, $-27^\circ$ and $-17^\circ$.
Correlating the lower-frequency \boom maps, which are 
dominated by CMB fluctuations, with the 3000~GHz 
map of \citet[model 8 of that paper]{finkbeiner99},
gave a measure of the 
spectral ratios between that map and the \boom bands;  these
ratios were used to scale the 410~GHz power spectra to the lower
frequencies.

\begin{figure}[htb]
\begin{center}
\resizebox{3in}{!}{
  \includegraphics[0.5in,2.0in][8in,10in]{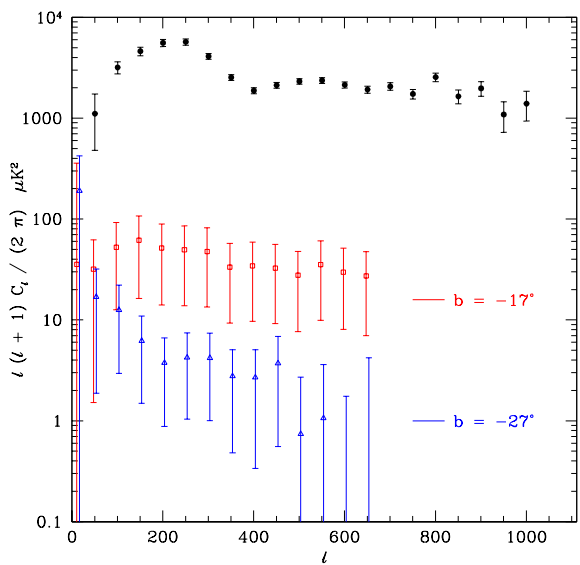}
}
\end{center}
\caption{\small
Angular power spectra of IRAS-correlated dust scaled to 150~GHz
for two circles of radius $9^\circ$ centered at galactic 
latitudes of $b=-17^\circ$ (red open squares) 
and $b=-27^\circ$ (blue open triangles).  Details of this
analysis can be found in \cite{masi01}.  The FASTER CMB power
spectrum (filled black circles) is shown for reference.
\label{fig:dust}
}
\end{figure}

In the region farthest from the galactic plane, the 410~GHz map is 
consistent with noise and no dust power spectrum result is reported.
Figure~\ref{fig:dust} shows the extracted power 
spectrum of dust for the $b=-27^\circ$ circle, taken directly from 
\cite{masi01}, along with the same calculation
for the $b=-17^\circ$ circle of that paper.  
These results show that the dust contribution to the total
power spectrum is largest at low $\ell$, and is generally
small ($< 100 \mu$K$^2$).

\begin{figure}[htb]
\begin{center}
\resizebox{3in}{!}{
  \includegraphics[0.5in,2.0in][8in,10in]{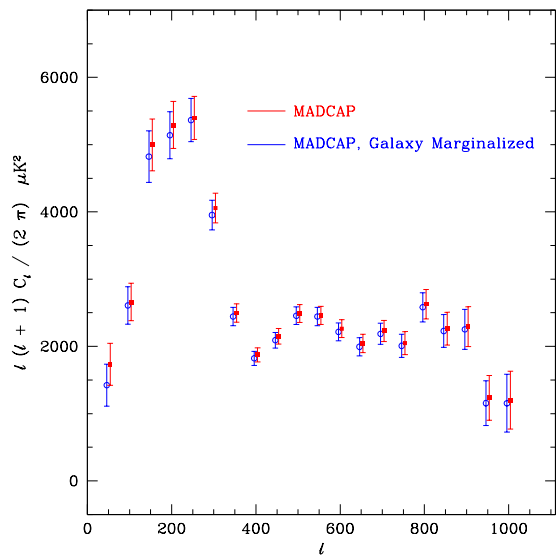}
}
\end{center}
\caption{\small
Galactic marginalization.  The filled red squares show the 
results of the MADCAP analysis with marginalization over the 
constant-declination 
modes (to remove constant-declination striping).  The 
open blue circles show the 
results after additional marginalization over two galactic templates, one of 
galactic dust and the other of galactic synchrotron emission.
These lead to slight shifts in the power spectrum at low $\ell$,
only significant in the first bin.
\label{fig:madcap_galmarg}
}
\end{figure}

A proper estimate of the contribution of dust emission to the
measured power spectrum requires the specific morphology of the
dust emission be taken
into account.  We have done this by using the MADCAP analysis path to
marginalize over templates of the galactic foregrounds.  The results
are shown in 
Figure~\ref{fig:madcap_galmarg}.  Here, we have used two templates,
one of galactic synchrotron emission~\citep{haslam81,rhodes98,finkbeiner01}, 
the other of galactic dust emission~\citep{schlegel98,finkbeiner01}.
The power spectrum is very stable to this process, with no significant
change for $\ell \ge 100$.  There is a one sigma change in 
the power at $\ell = 50$, consistent with the the 
expectation that the effects of dust
contamination should be largest at lowest $\ell$, and generally small.  We
use the galaxy template marginalized MADCAP results in the remainder
of this paper.  For the FASTER results, for which the statistical
errors at low $\ell$ are substantially higher than those of the MADCAP
spectrum, the effects of dust emission are negligible.

\section{Final Results}

\begin{figure}[htb]
\begin{center}
\resizebox{3in}{!}{
  \includegraphics[0.5in,2.0in][8in,10in]{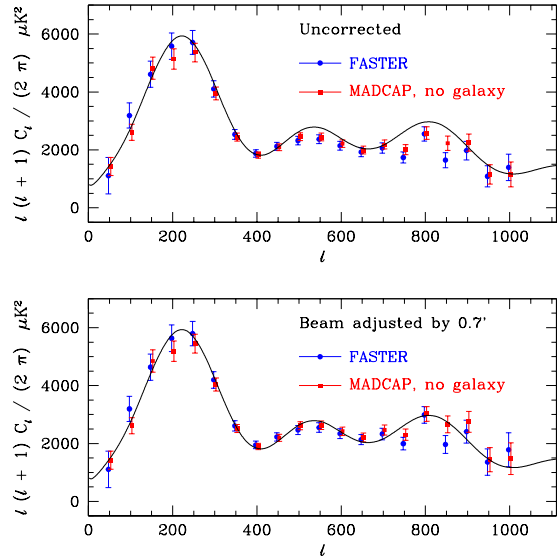}
}
\end{center}
\caption{\small
Final angular power spectrum results at 150~GHz, also given in 
Table~\ref{tab:APS}.  In both panels, the MADCAP results are shown 
as red squares, while the FASTER results are given as blue circles.
The top panel shows the data of Table~\ref{tab:APS}, along with 
the best-fit model from the weak-prior parameter estimation 
discussed below. 
In addition to the errors shown, there is a 10\% uncertainty in 
the temperature calibration (20\% in the temperature-squared units
of this plot), and a beam uncertainty of $1.4'$ rms.  In the bottom panel 
we have rescaled the data by changing the the beam window 
function by 0.5$\sigma$.  This gives much better
visual agreement with the model.
\label{fig:fullPS}
}
\end{figure}

We have used FASTER and MADCAP to derive two estimates of the 
angular power spectrum using the same input timestream, sky
coverage, and noise statistics.  The final FASTER results, derived
from a $3.5'$ pixel map and corrected for the small (1dps-2dps)/2
consistency test failure, appear along with the final galaxy-marginalized
MADCAP results in Figure~\ref{fig:fullPS}.  

Our power spectrum results are characterized by a likelihood function
for the bandpower in each band (${\cal C}_b$).  A good approximation
to this function is given by an offset lognormal
function~\citep{bond98} $Z_b = \ln(\overline{{\cal C}}_b + x_b)$ of the
maximum likelihood values found in each band ($\overline{{\cal C}}_b$)
and an offset parameter for each band, $x_b$.  Given these, the
likelihood is found by
\begin{eqnarray}
\sigma_b &=& \Delta \overline{{\cal C}}_b /  (\overline{{\cal C}}_b + x_b) \\
\Delta Z_b &=& \ln ({\cal C}_b + x_b) - \ln ( \overline{{\cal C}}_b + x_b) \\
-2 \ln L({\cal C}_b) &=& \sum_{bb'} \Delta Z_b  \sigma_b^{-1}  G_{bb'}
                \sigma_{b'}^{-1} \Delta Z_{b'} \label{bandlikelihood}
\end{eqnarray}
where
$G_{bb'}$ is the bandpower correlation matrix, normalized to 
unity on the diagonal.  Table~\ref{tab:APS} gives the maximum
likelihood value $\overline{{\cal C}}_b = \ell(\ell+1)C_\ell/2\pi$,
curvature error ($\Delta \overline{{\cal C}}_b$), and
offset parameter $x_b$ for each band for both the FASTER and
MADCAP results of Figure~\ref{fig:fullPS}.  The bin-bin
correlation matrices for these power spectra are given in
Tables~\ref{tab:fastermatrix} and \ref{tab:madcapmatrix}
for FASTER and MADCAP respectively.  These data, and the window 
functions of Figure~\ref{fig:windowfunctions}, are available
at \url{http://cmb.phys.cwru.edu/boomerang/} or 
\url{http://oberon.roma1.infn.it/boomerang}.

One measure of the level of agreement between the 
FASTER and MADCAP power spectra can be made by 
treating the two power spectra as independent datasets (which they 
are not) and using the curvature errorbars to calculate a chi-square
statistic.  We find that $\chi^2 = 8.54$ for 20 degrees of freedom,
which gives $P_> = 0.988$.  This low $\chi^2$ value indicates that the
two analyses of the same data vary by an amount much less than 
is expected for two random realizations of the same measurement.  
That is, the ``analysis variance" is very small compared to the 
statistical errors.

\begin{deluxetable}{ccccccccc}
\tablecolumns{9}
\tabletypesize{\scriptsize}
\tablecaption{Angular Power Spectra. \label{tab:APS}}
\tablewidth{0pt}
\tablehead{
\colhead{}&
\colhead{}&
\multicolumn{3}{c}{FASTER}&
\colhead{}&
\multicolumn{3}{c}{MADCAP}\\
\cline{3-5} \cline{7-9}\\
\colhead{$\ell_{low}$} & 
\colhead{$\ell_{high}$} &  
\colhead{$\overline{{\cal C}}_b$}   & 
\colhead{$\Delta \overline{{\cal C}}_b$}   & 
\colhead{$x_b$} & &
\colhead{$\overline{{\cal C}}_b$}   & 
\colhead{$\Delta \overline{{\cal C}}_b$}   &
\colhead{$x_b$} 
}
\startdata
 26 &   75 &  1053 & 401 &22 	&& 1423 & 313 &341\\
 76 &  125 &  3175 & 358 &40 	&& 2609 & 279 &34\\
126 &  175 &  4614 & 406 &71 	&& 4823 & 384 &50\\
176 &  225 &  5581 & 418 &110 	&& 5139 & 349 &81\\
226 &  275 &  5710 & 385 &162 	&& 5365 & 321 &124\\
276 &  325 &  4107 & 264 &228 	&& 3953 & 222 &180\\
326 &  375 &  2532 & 160 &320 	&& 2445 & 137 &249\\
376 &  425 &  1877 & 120 &441 	&& 1822 & 105 &337\\
426 &  475 &  2120 & 130 &593 	&& 2092 & 116 &467\\
476 &  525 &  2320 & 142 &794 	&& 2456 & 132 &638\\
526 &  575 &  2368 & 149 &1054 	&& 2444 & 135 &854\\
576 &  625 &  2141 & 147 &1397 	&& 2216 & 133 &1133\\
626 &  675 &  1923 & 149 &1838 	&& 1994 & 136 &1497\\
676 &  725 &  2066 & 170 &2437 	&& 2186 & 157 &2023\\
726 &  775 &  1738 & 184 &3202 	&& 2008 & 172 &2657\\
776 &  825 &  2551 & 239 &4204 	&& 2581 & 217 &3669\\
826 &  875 &  1647 & 252 &5542 	&& 2229 & 245 &4837\\
876 &  925 &  1976 & 312 &7237 	&& 2253 & 296 &6674\\
926 &  975 &  1087 & 352 &9696 	&& 1156 & 334 &8560\\
976 & 1025 &  1394 & 444 &12878 && 1155 & 430 &12324\\

\enddata

\tablecomments{\small
Angular power spectra of the CMB, derived using the 
FASTER (columns 3-5) and MADCAP (columns 6-8) methods.
The FASTER power spectrum has been corrected for the 
(1-2dps)/2 failure by the addition of a systematic errorbar
in quadrature with the statistical one in the relevant 
$\ell$ bins.  The MADCAP power spectrum has been marginalized 
over two galactic templates as discussed in the text.
The FASTER power spectrum is calculated for shaped bins, while
the MADCAP power spectrum is calculated for tophat bins.
}

\end{deluxetable}

\begin{deluxetable}{cccccccccccccccccccc}
\rotate
\tablecolumns{20}
\tabletypesize{\tiny}
\tablecaption{FASTER Bandpower Correlation Matrix\label{tab:fastermatrix}}
\tablewidth{0pt}
\tablehead{
\colhead{1}&
\colhead{2}&
\colhead{3}&
\colhead{4}&
\colhead{5}&
\colhead{6}&
\colhead{7}&
\colhead{8}&
\colhead{9}&
\colhead{10}&
\colhead{11}&
\colhead{12}&
\colhead{13}&
\colhead{14}&
\colhead{15}&
\colhead{16}&
\colhead{17}&
\colhead{18}&
\colhead{19}&
\colhead{20}
}
\startdata
 1.000 &-0.140 & 0.005 &-0.002 & 0 & 0 & 0 & 0 & 0 & 0 & 0 & 0 & 0 & 0 & 0 & 0 & 0 & 0 & 0 & 0  \\
 - & 1.000 &-0.089 & 0 &-0.004 &-0.004 &-0.005 &-0.007 &-0.006 &-0.005 &-0.006 &-0.006 &-0.007 &-0.007 &-0.007 &-0.006 &-0.006 &-0.005
&-0.005 &-0.004  \\
 - & - & 1.000 &-0.088 &-0.001 &-0.006 &-0.006 &-0.007 &-0.007 &-0.006 &-0.005 &-0.006 &-0.006 &-0.006 &-0.007 &-0.005 &-0.006 &-0.005
&-0.005 &-0.004  \\
 - & - & - & 1.000 &-0.087 &-0.003 &-0.008 &-0.008 &-0.006 &-0.006 &-0.005 &-0.005 &-0.006 &-0.005 &-0.005 &-0.005 &-0.005 &-0.004 &-0.004 &-0.004  \\
 - & - & - & - & 1.000 &-0.088 &-0.005 &-0.009 &-0.006 &-0.004 &-0.004 &-0.004 &-0.004 &-0.004 &-0.004 &-0.003 &-0.004 &-0.003 &-0.003
&-0.003  \\
 - & - & - & - & - & 1.000 &-0.090 &-0.004 &-0.005 &-0.003 &-0.003 &-0.003 &-0.002 &-0.002 &-0.002 &-0.002 &-0.002 &-0.002 &-0.002 &-0.002  \\
 - & - & - & - & - & - & 1.000 &-0.088 & 0 &-0.003 &-0.002 &-0.001 &-0.001 &-0.001 &-0.001 & 0 & 0 & 0 & 0 & 0  \\
 - & - & - & - & - & - & - & 1.000 &-0.085 & 0 &-0.003 &-0.002 &-0.001 & 0 & 0 & 0 & 0 & 0 & 0 & 0  \\
 - & - & - & - & - & - & - & - & 1.000 &-0.088 & 0 &-0.003 &-0.002 &-0.001 &-0.001 & 0 & 0 & 0 & 0 & 0  \\
 - & - & - & - & - & - & - & - & - & 1.000 &-0.089 & 0 &-0.003 &-0.002 &-0.001 & 0 & 0 & 0 & 0 & 0  \\
 - & - & - & - & - & - & - & - & - & - & 1.000 &-0.089 & 0 &-0.003 &-0.002 & 0 & 0 & 0 & 0 & 0  \\
 - & - & - & - & - & - & - & - & - & - & - & 1.000 &-0.087 & 0 &-0.003 &-0.001 & 0 & 0 & 0 & 0  \\
 - & - & - & - & - & - & - & - & - & - & - & - & 1.000 &-0.084 & 0 &-0.003 &-0.001 & 0 & 0 & 0  \\
 - & - & - & - & - & - & - & - & - & - & - & - & - & 1.000 &-0.086 & 0 &-0.003 &-0.001 & 0 & 0  \\
 - & - & - & - & - & - & - & - & - & - & - & - & - & - & 1.000 &-0.088 & 0.001 &-0.003 &-0.001 & 0  \\
 - & - & - & - & - & - & - & - & - & - & - & - & - & - & - & 1.000 &-0.091 & 0 &-0.003 &-0.001  \\
 - & - & - & - & - & - & - & - & - & - & - & - & - & - & - & - & 1.000 &-0.088 & 0 &-0.003  \\
 - & - & - & - & - & - & - & - & - & - & - & - & - & - & - & - & - & 1.000 &-0.087 & 0  \\
 - & - & - & - & - & - & - & - & - & - & - & - & - & - & - & - & - & - & 1.000 &-0.081  \\
 - & - & - & - & - & - & - & - & - & - & - & - & - & - & - & - & - & - & - & 1.000  
\enddata
\tablecomments{\small
The FASTER CMB power spectrum band-band correlation matrix, $G_{bb'}$
of equation~\ref{bandlikelihood}.
This matrix is symmetric;  values below the diagonal, not
printed, are symmetric with those above.
Values with magnitude $0.0005$ and lower have been truncated
to zero.  Bands are labelled 1-20 in consecutive order from
low to high $\ell$, as given in Table~\ref{tab:APS}.
}
\end{deluxetable}

\begin{deluxetable}{cccccccccccccccccccc}
\rotate
\tablecolumns{20}
\tabletypesize{\tiny}
\tablecaption{MADCAP Bandpower Correlation Matrix\label{tab:madcapmatrix}}
\tablewidth{0pt}
\tablehead{
\colhead{1}&
\colhead{2}&
\colhead{3}&
\colhead{4}&
\colhead{5}&
\colhead{6}&
\colhead{7}&
\colhead{8}&
\colhead{9}&
\colhead{10}&
\colhead{11}&
\colhead{12}&
\colhead{13}&
\colhead{14}&
\colhead{15}&
\colhead{16}&
\colhead{17}&
\colhead{18}&
\colhead{19}&
\colhead{20}
}
\startdata
 1.000 &-0.080 &-0.001 &-0.002 & 0 & 0 & 0 & 0 & 0 & 0 & 0 & 0 & 0 & 0 & 0 & 0 & 0 & 0 & 0 & 0  \\
 - & 1.000 &-0.058 &-0.001 &-0.001 & 0 & 0 & 0 & 0 & 0 & 0 & 0 & 0 & 0 & 0 & 0 & 0 & 0 & 0 & 0  \\
 - & - & 1.000 &-0.057 &-0.001 &-0.001 & 0 & 0 & 0 & 0 & 0 & 0 & 0 & 0 & 0 & 0 & 0 & 0 & 0 & 0  \\
 - & - & - & 1.000 &-0.056 & 0 &-0.001 & 0 & 0 & 0 & 0 & 0 & 0 & 0 & 0 & 0 & 0 & 0 & 0 & 0  \\
 - & - & - & - & 1.000 &-0.055 & 0 &-0.001 & 0 & 0 & 0 & 0 & 0 & 0 & 0 & 0 & 0 & 0 & 0 & 0  \\
 - & - & - & - & - & 1.000 &-0.055 & 0 &-0.001 & 0 & 0 & 0 & 0 & 0 & 0 & 0 & 0 & 0 & 0 & 0  \\
 - & - & - & - & - & - & 1.000 &-0.056 & 0 &-0.001 & 0 & 0 & 0 & 0 & 0 & 0 & 0 & 0 & 0 & 0  \\
 - & - & - & - & - & - & - & 1.000 &-0.055 & 0 &-0.002 & 0 & 0 & 0 & 0 & 0 & 0 & 0 & 0 & 0  \\
 - & - & - & - & - & - & - & - & 1.000 &-0.054 & 0 &-0.002 & 0 & 0 & 0 & 0 & 0 & 0 & 0 & 0  \\
 - & - & - & - & - & - & - & - & - & 1.000 &-0.054 & 0 &-0.002 & 0 & 0 & 0 & 0 & 0 & 0 & 0  \\
 - & - & - & - & - & - & - & - & - & - & 1.000 &-0.054 &-0.001 &-0.002 & 0 & 0 & 0 & 0 & 0 & 0  \\
 - & - & - & - & - & - & - & - & - & - & - & 1.000 &-0.055 &-0.001 &-0.002 & 0 & 0 & 0 & 0 & 0  \\
 - & - & - & - & - & - & - & - & - & - & - & - & 1.000 &-0.055 &-0.002 &-0.002 &-0.001 & 0 & 0 & 0  \\
 - & - & - & - & - & - & - & - & - & - & - & - & - & 1.000 &-0.056 &-0.002 &-0.002 &-0.001 & 0 & 0  \\
 - & - & - & - & - & - & - & - & - & - & - & - & - & - & 1.000 &-0.056 &-0.002 &-0.002 &-0.001 & 0  \\
 - & - & - & - & - & - & - & - & - & - & - & - & - & - & - & 1.000 &-0.057 &-0.002 &-0.002 &-0.001  \\
 - & - & - & - & - & - & - & - & - & - & - & - & - & - & - & - & 1.000 &-0.058 &-0.002 &-0.002  \\
 - & - & - & - & - & - & - & - & - & - & - & - & - & - & - & - & - & 1.000 &-0.060 &-0.003  \\
 - & - & - & - & - & - & - & - & - & - & - & - & - & - & - & - & - & - & 1.000 &-0.061  \\
 - & - & - & - & - & - & - & - & - & - & - & - & - & - & - & - & - & - & - & 1.000  \\
\enddata
\tablecomments{\small
The MADCAP CMB power spectrum band-band correlation matrix, $G_{bb'}$
of equation~\ref{bandlikelihood}.
This matrix is symmetric;  values below the diagonal, not
printed, are symmetric with those above.
Values with magnitude $0.0005$ and lower have been truncated
to zero.  Bands are labelled 1-20 in consecutive order from
low to high $\ell$, as given in Table~\ref{tab:APS}.
}
\end{deluxetable}

\section{Features in the Power Spectrum}

The cosmological parameter estimation procedure we follow below
is done in the context of inflation-motivated models with
adiabatic initial density perturbations.  Thus, it is both
interesting and important to assess the evidence
in favor of these models.  One of their generic predictions is
that there will be a series of peaks in the CMB
power spectrum, the exact positions and amplitudes of which depend
on the cosmological parameters.  It is thus interesting to
search for such features in our power spectrum and evaluate
the statistical significance with which they are detected.

To detect such features, we use the method applied
to the B02 power spectrum in \cite{debernardis01}, based on parabolic
fits to the CMB power spectrum over a fixed number of bands.  We fit
the spectrum to the polynomial
\begin{equation}
  {\cal C}_{\ell} = {\cal C}_A (\ell-\ell_p)^2 + {\cal C}_B,
\end{equation}
where $\ell_p$ is the peak position.  In order to fit the measured
bandpowers $\overline{\cal C}_b$ we average the model ${\cal C}_\ell$ over
the bands reported
in Table~\ref{tab:APS}, thus obtaining the 
theoretical bandpowers ${\cal C}_{b}^T$.
Using the covariance matrix $G^{-1}_{bb^{'}}$ of the measured
bandpowers we compute
\begin{equation}
  \chi^2= (\overline{{\cal C}}_b-{\cal C}_{b}^T) G^{-1}_{bb'}
(\overline{{\cal C}}_{b'} - {\cal C}_{b'}^T),
\end{equation}
which we minimize by varying ${\cal C}_A$, ${\cal C}_B$ 
and $\ell_p$.
Errors on the fit $\ell_p$ and ${\cal C}_B$ are found by marginalization
of the full likelihood over ${\cal C}_A$.
In order to evaluate the significance of the detection of a
feature we study the
likelihood of the curvature ${\cal C}_A$ marginalizing over the other two
parameters.

\begin{figure}[htb]
 \begin{center}
 \resizebox{3.0in}{!}{
   \includegraphics{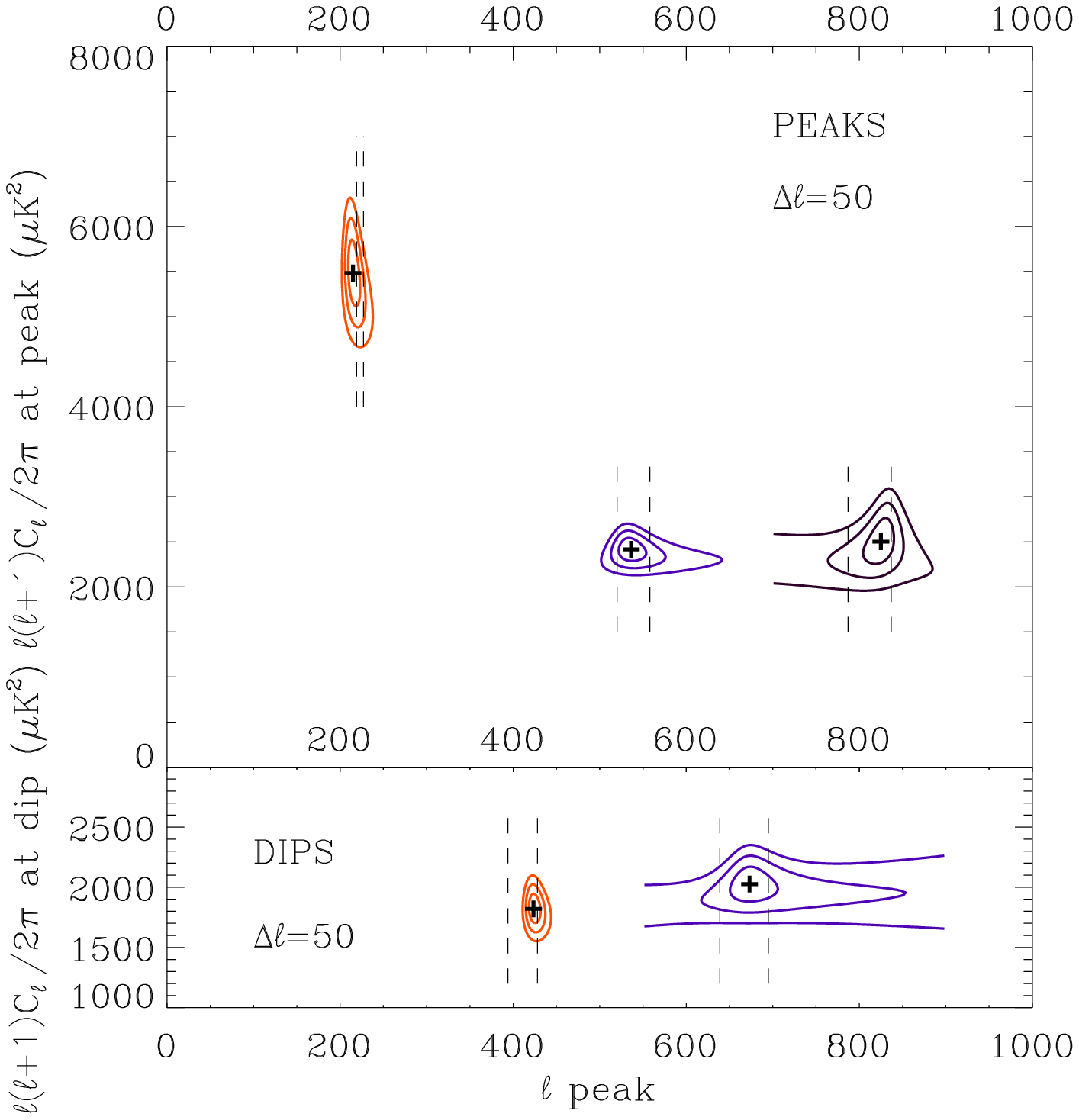}
 }
 \end{center}
\caption{$\Delta \chi^2$ contours for the position and amplitude of
peaks and valleys in the \boom MADCAP power spectrum.  
The contours are at $\Delta \chi^2 =2.3,6.17,11.8$, 
i.e. 68.3\%, 95.4\% and 99.7\% confidence.  The vertical dashed
lines give the feature positions found using
the cosmological database, and as given in Table~\ref{tab:peaks}.
Similar results are obtained using the FASTER results. 
\label{fig:peaks}
}
\end{figure}

When we compare different models, i.e. different values of the two
parameters $\ell_p$ and ${\cal C}_B$, the
$\chi^2$ has 2 degrees of freedom.  In order to show how other models
compare to the best-fit one, we plot in
Figure~\ref{fig:peaks} the contours corresponding to $\Delta \chi^2
=2.3,6.17,11.8$, i.e. $68.3\%, 95.4\%$ and $99.7\%$ confidence.

Table~\ref{tab:peaks} shows the results of this analysis for both
the MADCAP and the FASTER power spectra of Table~\ref{tab:APS}.
The significance of the detections depends
somewhat on the range of bands over which the fit is done;
the results in the table are those that give the most significant
detections.
Comparing the results to \cite{debernardis01} we find a general
improvement in the precision with which the peaks and valleys are
located, particularly for the first and second peaks, and for
the first valley.
We obtain very similar results in a variation of this method where
a three-parameter quadratic is fit over a 
sliding five-band window, also described in \cite{debernardis01}. 
The results are also very similar when applied to a FASTER 
power spectrum derived for bands of the same width ($\Delta \ell = 50$)
with band centers shifted by $\ell = 25$.

In order to investigate the level at which the detections 
of different peaks are correlated, we have performed a 
simultaneous fit of all the spectral bins using a linear 
combination of four Gaussians
\begin{equation}
  {\cal C}_\ell = 
  	\sum_{i=1}^4 A_i^2 \exp (-(\ell-\ell_i)^2/2\sigma_i^2),
\end{equation}
which is sufficiently flexible to provide a good 
fit to any standard theoretical spectrum.  We proceed using 
a Monte-Carlo Markov Chain method, as 
in \citet{christensen01}, \citet{lewis02}, and \citet{odman02}, 
accounting for calibration and beam uncertainties as in \citet{bridle02}.
We find best fit values for $A_i$ and $\ell_i$ that are in good agreement
with the results obtained above, and point clearly
towards the presence of features in the power spectrum.

Using this simultaneous fit to all of the power spectrum bins with 
a single phenomenological function allows us to study the 
correlation between the different parameters.  These are not 
negligible between the amplitudes of the peaks that are near to 
each other (for example $R(A_1,A_2)=0.19$; $R(A_1,A_3)=0.07$, 
$R(A_2,A_3)=0.27$), and between amplitudes and widths
($R(A_1,\sigma_1)=0.20$), but the detections are all confirmed.

As the table and figure show, we clearly detect 
multiple features in the power spectrum.  The next question 
is whether the adiabatic perturbation, inflationary model
set can produce models with similar features.

Using the same methods discussed below
for cosmological parameter estimation, we use the data and our 
theoretically motivated database of ${\cal C}_\ell$ models 
to make Bayesian estimates of the 
positions and amplitudes of peaks in the power spectrum, for 
comparison with our model-independent fits.
The last two columns of Table~\ref{tab:peaks} 
show the results of this process (using the ``weak prior" described
below), and give 
results that agree
very well with the phenomenologically measured parameters of
the various features.  This bolsters our confidence in the
model set we use in the next section, to estimate cosmological
parameters.

\begin{deluxetable}{cccccccccccccc}
\tablecaption{Peaks and Valleys in the CMB \label{tab:peaks}}
\tablecolumns{11}
\tabletypesize{\scriptsize}
\tablewidth{0pt}
\tablehead{
\colhead{}&
\colhead{}&
\colhead{}&
\multicolumn{3}{c}{MADCAP}&
\colhead{}&
\multicolumn{3}{c}{FASTER}&
\colhead{}&
\multicolumn{2}{c}{Adiabatic CDM}&
\colhead{}\\
\cline{4-6} \cline{8-10} \cline{12-13} \\
\colhead{Feature} &
\colhead{$\ell$ range}&
\colhead{}&
\colhead{$\ell_{\rm p}$} &
\colhead{${\cal C}_{\rm p}$ ($\mu K^{2}$)}   &
\colhead{level}   &
\colhead{}&
\colhead{$\ell_{\rm p}$} &
\colhead{${\cal C}_{\rm p}$ ($\mu K^{2}$)}   &
\colhead{level} &
\colhead{}&
\colhead{$\ell_{\rm p}$} &
\colhead{${\cal C}_{\rm p}$ ($\mu K^{2}$)}  \\
}
\startdata
Peak 1 & $100-300$ & & $216^{+6}_{-5}$ & $5480^{+1130}_{-1130}$ 
& $6.7\sigma$ & &
$215^{+5}_{-6}$  & $5690^{+1200}_{-1200}$ & $4.8\sigma$  & &
$223^{+4}_{-4}$  & $6022^{+394}_{-370}$\\

Valley 1 & $300-500$ & & $425^{+4}_{-5}$ & $1820^{+420}_{-410}$   
& $6.3\sigma$ & &
$430^{+7}_{-5}$& $1870^{+420}_{-410}$ & $4.3\sigma$ & &
$411^{+17}_{-17}$ & $1881^{+152}_{-141}$  \\

Peak 2 & $400-650$ & & $536^{+10}_{-10}$ & $2420^{+620}_{-570}$ 
& $4.0\sigma$ & &
$528^{+14}_{-10}$ & $2330^{+600}_{-550}$ & $3.1\sigma$  & &
$539^{+19}_{-19}$  & $2902^{+248}_{-229}$ \\

Valley 2 & $550-800$ & & $673^{+18}_{-13}$ & $2030^{+670}_{-560}$ 
& $2.6\sigma$ & &
$681^{+21}_{-21}$ & $1910^{+630}_{-530}$ & $2.5\sigma$  & &
$667^{+28}_{-27}$ & $2122^{+302}_{-265}$  \\

Peak 3 & $750-950$ & & $825^{+10}_{-13}$ & $2500^{+1100}_{-840}$ 
& $3.2\sigma$ & &
$820^{+15}_{-22}$ & $2150^{+1000}_{-720}$ & $2.2\sigma$ & &
$812^{+26}_{-25}$ & $3121^{+497}_{-429}$ \\
\enddata
\tablecomments{\small
Locations and amplitudes of peaks and valleys in the
power spectrum of the CMB, obtained with polynomial fits.
The locations, amplitudes, and confidence levels of
detection are listed for MADCAP (columns 3-5) and
FASTER (columns 6-8).  The $\ell$ range used in the parabolic 
analysis is reported in column 2.  Columns 9 and 10 gives the result
of cosmological ``peak parameter" extraction (using the
MADCAP data, COBE-DMR data, and the ``weak cosmological 
prior" discussed below) from the
set of adiabatic perturbation, cold dark matter models
used in our cosmological parameter estimation.  
All the errors include the effects of gain and beam 
calibration uncertainties.}
\end{deluxetable}

\section{Cosmological Parameters}

Our measurement of the CMB angular power spectrum can 
be used in conjunction with other cosmological information
to constrain several cosmological parameters.  Our method,
described in detail in \cite{lange01}, compares the measured
angular power spectrum with the predicted power spectra from a
family of theoretical models.  We choose to compare our measurements
with inflation-motivated adiabatic cold dark matter 
models, with the 7 cosmological parameters given 
in Table~\ref{tab:params}.

We take a Bayesian approach, calculating a likelihood of each model
given the data, in the discrete parameter database of
Table~\ref{tab:params}.  We then marginalize over
the continuous parameters such as theory 
normalization ($\ln {\cal C}_{10}$), 
calibration, and beam uncertainty for each model.  To find 
confidence intervals on 
any given parameter, we marginalize over the other parameters by 
integrating through the database, collapsing the n-dimensional likelihood
to a one-dimensional likelihood curve for that parameter.

In the comparison of the theoretical and measured power spectra, one
must convolve the predicted theory power spectrum with the window
function for each $\ell$ bin of the measurement.  The flat band average of a
target model ${\cal C}^T_{\ell} = \ell(\ell+1)C^T_{\ell}/2\pi$, can be
defined with respect to a window function $W^b_{\ell}$ for that 
band as
\begin{equation}
{\cal C}^T_b = \frac{{\cal I}[ W^b_{\ell} {\cal C}^T_{\ell}]}
		    {{\cal I}[W^b_{\ell}]}
\end{equation}
with
\begin{equation}
   {\cal I}[f_{\ell}] =
     \sum_{\ell} \frac{(\ell+\frac{1}{2})}
     		      {\ell(\ell+1)}       
		      f_{\ell} .
\end{equation}

In the power spectrum estimation pipelines discussed above, we can
choose to use shaped bands rather than flat.  This will change
the details of the window function, but the prescription for
calculating theoretical band averages remains the same.

\begin{figure}[htb]
\begin{center}
\resizebox{3in}{!}{
  \rotatebox{-90}{
  \includegraphics[0.5in,0in][8in,11in]{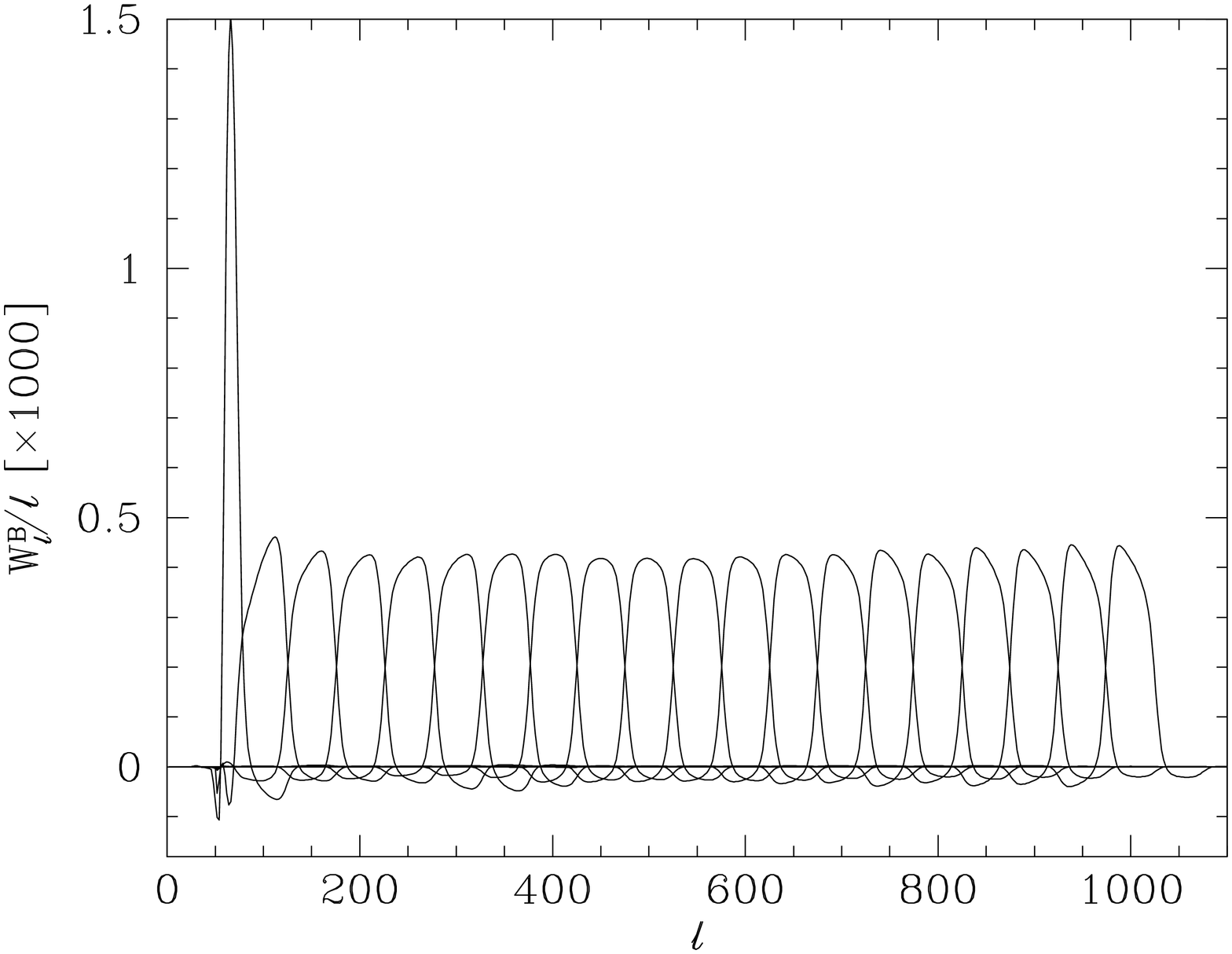}
  }
}
\end{center}
\caption{\small
Window functions derived from the FASTER analysis. 
These window functions are used to relate a
continuous theoretical model to the expected experimental band powers, a
crucial step in parameter extraction.  The functions are
orthogonal, as a result of the top-hat binning assumed in the theory.
The window for the first band shows how all
the information is coming from the $50<\ell<76$ region due to the sharp
filtering applied (in the FASTER pipeline) to the timestream. 
\label{fig:windowfunctions}
}
\end{figure}

We have calculated the window functions for the FASTER power spectrum
bins, using S+N weighting on the map.  In 
Figure~\ref{fig:windowfunctions} we show the
flat-band window functions, to illustrate their $\ell$-space
shapes and the level of correlations between bands.  
Details on their derivation are given in
\cite{contaldi02}.  For the MADCAP comparison with theory, we 
use tophat window functions.

\begin{deluxetable}{cl}
\tabletypesize{\scriptsize}
\tablecaption{Cosmological Parameters Database \label{tab:params}}
\tablewidth{0pt}
\tablehead{
\colhead{Parameter} & 
\colhead{Values}  
}
\startdata
$\Omega_{k}$    & -0.5, -0.3, -0.2, -0.15, -0.1, -0.05, 0, 0.05, 0.1, 0.15, 
		0.2, 0.3, 0.5, 0.7, 0.9 \\[0.5ex] \\
$\Omega_\Lambda$ & 0, 0.1, 0.2, 0.3, 0.4, 0.5, 0.6, 0.7,
                   0.8, 0.9, 1.0, 1.1 \\  \\
$\omega_c $     & 0.03, 0.06, 0.08, 0.10, 0.12, 0.14, 0.17, 0.22, 
			0.27, 0.33, 0.40, 0.55, 0.8 \\[0.5ex]  \\
$\omega_b $     & 0.003125, 0.00625, 0.0125, 0.0175, 0.020, 0.0225, 
		  0.025, 0.030, 0.035,\\ 
		& 0.04, 0.05, 0.075, 0.10, 0.15, 0.2 \\[0.5ex]  \\
$n_s$           & 0.5, 0.55, 0.6, 0.65, 0.7, 0.725, 0.75, 0.775, 0.8,
		0.825, 0.85, 0.875, 0.9, \\
		& 0.925, 0.95, 0.975, 1.0,
		1.025, 1.05, 1.075, 1.1, 1.125, 1.15, 1.175, 1.2,\\
		&1.25, 1.3, 1.35, 1.4, 1.45, 1.5 \\[0.5ex]  \\
$\tau_c$        & 0, 0.025, 0.05, 0.075, 0.1, 0.15, 0.2, 0.3, 0.4, 
			0.5, 0.7 \\[0.5ex]  \\
$\ln{{\cal }C_{10}}$   & Continuous \\[0.5ex] 

\enddata

\tablecomments{\small
The values of the cosmological parameters in our model space;
while $\ln{C_{10}}$ is marginalized as a continuous variable,
the rest are calculated on a grid with the discrete parameter
values given.  The curvature $\Omega_{k}$ is related to the
overall density by $\Omega_{k} = 1 - \Omega_{total}$.  The cold
dark matter and baryon physical densities $\omega_c$ and
$\omega_b$ are given by $\omega_x = \Omega_{x} h^2$, where
$h$ is the Hubble parameter in units of 100~km/s/MPc.
The database is restricted to models for which 
$\Omega_{M}$ = $\Omega_c + \Omega_b >$ 0.1.
$n_s$ is the spectral index for primordial density fluctuations,
where a value of 1.0 indicates scale invariance.  Reionization
is parameterized by $\tau_c$, the Thompson depth to the
epoch when the universe reionized after photon decoupling.  
In addition to these cosmological parameters, there are 
instrumental parameters
describing the systematic gain and beam uncertainties.  These are
accounted for, by marginalization, in all cosmological parameter 
estimates reported in this paper.
}

\end{deluxetable}

We can also apply a series of ``priors", or prior probabilities, to each 
model in the database, modifying the likelihood of that model before
marginalization.  The priors we choose include a ``weak prior"
which sets the likelihood to zero if, for that model, the Hubble 
parameter ($H_o = 100 h$km/s/MPc) has a value outside
the range $0.45 < h < 0.90$, the current 
age of the universe is less than 10~Gyr, or the total matter
content $\Omega_M < 0.1$.  We also investigate the
effect of narrowing the prior on the Hubble constant to 
$h = 0.72 \pm 0.08$, as measured by the Hubble Key Project~\citep{freedman01}.

We also examine the effect of a large-scale structure 
(LSS) prior.  This is a joint 
constraint on $\sigma_8^2$, the bandpower of (linear) density fluctuations 
on a scale corresponding to rich clusters of galaxies ($8 h^{-1}$MPc), and on a
shape parameter $\Gamma_{\rm eff}$ characterizing the (linear) density
power spectrum.  The LSS prior probability distribution, described in
detail in \citet{bond02}, is slightly modified over that used in
\citet{lange01} to agree better with weak lensing and clustering
data.  $\sigma_8 \Omega_m^{0.56}$ = $ 0.47^{+.02,+.11}_{-.02,-.08}$, is
distributed as a Gaussian (first error) convolved with a uniform (top-hat)
distribution (second error), centered about 0.47; $\Gamma_{\rm eff}$ =
$0.21^{+.03,+.08}_{-.03,-.08}$ is a broad distribution over the 0.1 to 
0.3 range.
Here $\Gamma_{\rm eff}$ = $\Gamma + (n_s - 1)/2$, where
$\Gamma \approx \Omega_m h \, \exp[-\Omega_B(1+\Omega_m^{-1}(2{
h})^{1/2})]$ is a function of our basic cosmological parameters.

Our final set of priors combines the weak and LSS priors with the
supernova data of \citet{riess98} and \citet{perlmutter99b}, 
and the assumption that the geometry of space is flat.  
In all cases, we use the COBE-DMR measurements~\citep{bennett96} 
to provide a valuable low-$\ell$ anchor for the power spectrum.

\begin{figure}[htb]
\begin{center}
\resizebox{3.5in}{!}{
  \includegraphics[0.5in,2in][8.5in,9in]{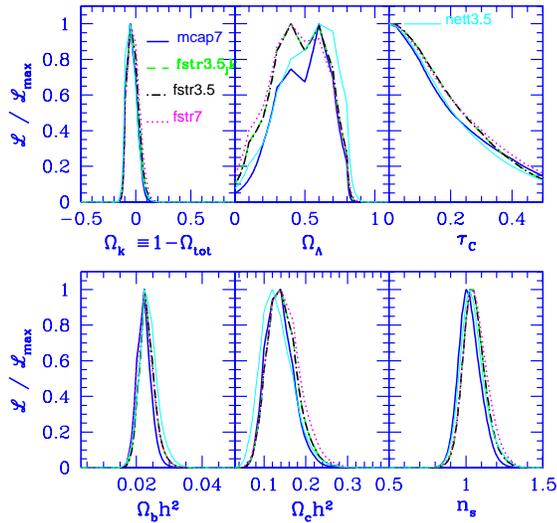}
  }
\end{center}
\caption{\small
Parameters extracted from our data plus the COBE-DMR 
results~\citep{bennett96}, using two ``weak" priors, $0.4<h<0.9$
and age~$> 10$~Gyr.  These likelihood curves are quite insensitive to
variations in the method or details of the analysis, and show that
our analysis methods are quite robust.
The green dashed curve shows the FASTER results 
of Table~\ref{tab:APS}, while the solid blue shows the 
MADCAP results of that same table.  The black dot-dash curve shows
the FASTER result with no correction for the 
(1dps-2dps)/2 consistency test failure.
The magenta dotted curve shows the results upon degrading the 
FASTER analysis to $7'$ resolution.  The cyan solid curve shows
the (FASTER derived) B02 results, using less timestream data 
and sky coverage as discussed in the text.  
\label{fig:params_weaks}
}
\end{figure}

We are interested in the robustness of our parameter
extraction to the details of the input power spectrum.  Specifically,
we would like to know if the small differences between different
variations of the FASTER analysis, or between the final FASTER and 
MADCAP power spectra, lead to significant differences in 
cosmological results.  
In Figure~\ref{fig:params_weaks} we show likelihood curves for six
cosmological parameters derived using only the weak prior case, 
for several input versions of our angular power spectrum results.
In all cases the likelihood curves are very similar, indicating 
the cosmological results are not very sensitive to the details
of our analysis.

\begin{figure}[htb]
\begin{center}
\resizebox{3.5in}{!}{
  \includegraphics[0.5in,2in][8.5in,9in]{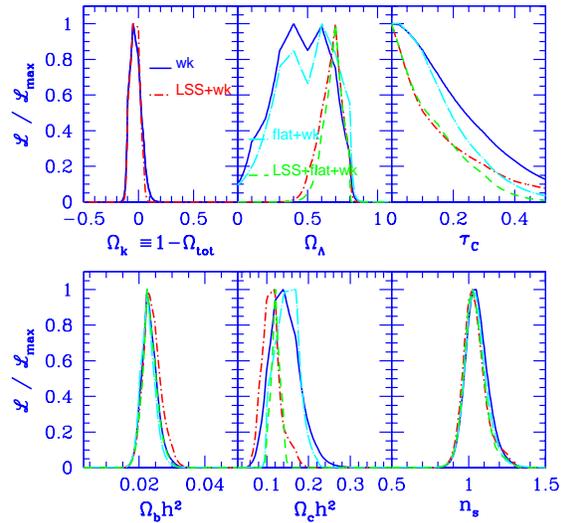}
  }
\end{center}
\caption{\small
Likelihood curves for six cosmological parameters, derived
from the FASTER power spectrum of Table~\ref{tab:APS}
and COBE-DMR, for a series of applied priors described
in the text.  The solid blue curve is for the ``weak prior" 
case.  The dot-dash red line adds the LSS prior to the 
weak prior.  The cyan dot-dash curve is for the ``weak prior"
case with the added assumption that the geometry is flat.  
The green dashed curve adds to this the LSS prior.
Three parameters, $\Omega_k$, $\Omega_b h^2$, and $n_s$ are 
very well determined and unaffected by the choice of prior.
$\Omega_c h^2$ is fairly well localized by the weak and 
weak+flat cases, but much better determined when an LSS prior is
applied.  Similarly, the limits on $\tau_c$ improve with the 
use of the LSS prior.  The data favor a non-zero 
$\Omega_\Lambda$ in the weak and weak+flat cases; the use
of an LSS prior leads to a very solid detection.  In all
cases, the observed parameters are consistent with 
a flat, $\Lambda$-CDM cosmology.
\label{fig:params_faster}
}
\end{figure}

\begin{figure}[htb]
\begin{center}
\resizebox{3.5in}{!}{
 \includegraphics[0.5in,2in][8.5in,9in]{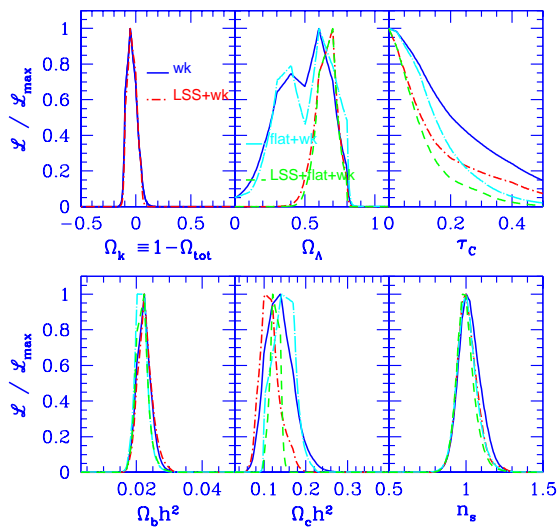}
 }
\end{center}
\caption{\small
Likelihood curves for six cosmological parameters, derived
from the MADCAP power spectrum of Table~\ref{tab:APS}
and COBE-DMR, for a series of applied priors described
in the text.  The curves are as in Figure~\ref{fig:params_faster},
and lead to the same conclusions.  The agreement of the curves in this
figure with those in Figure~\ref{fig:params_faster} demonstrates
the insensitivity of our analysis to the details of the CMB angular power
spectrum estimation pipeline.
\label{fig:params_madcap}
}
\end{figure}

Having demonstrated the stability of our results, 
we now turn to extracting cosmological parameters from 
our angular power spectrum with the series of applied priors
discussed above.
Figures~\ref{fig:params_faster} and 
Figures~\ref{fig:params_madcap} show a 
set of one dimensional likelihood curves for six 
parameters, derived from the data of Table~\ref{tab:APS}, 
COBE-DMR, and the priors described above.  
Inspection of these figures shows that the 
parameter likelihoods derived from the FASTER and MADCAP results 
are very similar for each set of priors.  This again demonstrates
the stability of the cosmological results to the 
chosen analysis path.  Numerical estimates of 
parameters derived from these curves are given 
in Table~\ref{tab:params}, where a similar comparison can be made.

%
\begin{deluxetable}{llllllllllll}
\tablecolumns{12}
\tabletypesize{\scriptsize}
\tablecaption{Parameter Estimates}
\tablewidth{0pt}
\tablehead{
\colhead{Priors} 
&\colhead{Analysis Path}
& \colhead{$\Omega_{tot}$}
& \colhead{$n_s$}
& \colhead{$\Omega_bh^2$}
& \colhead{$\Omega_{cdm}h^2$}
& \colhead{$\Omega_{\Lambda}$}
& \colhead{$\Omega_m$}
& \colhead{$\Omega_b$}
& \colhead{$h$}
& \colhead{Age}
& \colhead{$\tau_c$}
}
\startdata
\multicolumn{2}{l}{Weak $h$ + age} \\
& MADCAP
& $1.04^{0.05}_{0.06}$
& $1.02^{0.08}_{0.07}$
& $0.023^{0.003}_{0.003}$
& $0.14^{0.04}_{0.04}$
& $0.50^{0.18}_{0.21}$
& $0.55^{0.19}_{0.19}$
& $0.077^{0.023}_{0.023}$
& $0.56^{0.10}_{0.10}$
& $14.8^{1.4}_{1.4}$
& $<0.52$
\\
& FASTER
& $1.03^{0.05}_{0.06}$
& $1.05^{0.08}_{0.07}$
& $0.023^{0.003}_{0.003}$
& $0.14^{0.04}_{0.04}$
& $0.46^{0.21}_{0.21}$
& $0.58^{0.21}_{0.21}$
& $0.080^{0.025}_{0.025}$
& $0.56^{0.11}_{0.11}$
& $14.5^{1.5}_{1.5}$
& $<0.51$
\\
\multicolumn{2}{l}{Weak $h$ + age + LSS} \\
& MADCAP
& $1.03^{0.05}_{0.05}$
& $1.01^{0.07}_{0.06}$
& $0.023^{0.003}_{0.003}$
& $0.11^{0.03}_{0.02}$
& $0.66^{0.07}_{0.09}$
& $0.38^{0.10}_{0.10}$
& $0.065^{0.020}_{0.020}$
& $0.61^{0.11}_{0.11}$
& $14.9^{1.7}_{1.7}$
& $<0.50$
\\
& FASTER
& $1.03^{0.05}_{0.05}$
& $1.03^{0.08}_{0.07}$
& $0.024^{0.004}_{0.003}$
& $0.11^{0.03}_{0.03}$
& $0.68^{0.07}_{0.10}$
& $0.36^{0.11}_{0.11}$
& $0.066^{0.022}_{0.022}$
& $0.63^{0.11}_{0.11}$
& $14.8^{1.7}_{1.7}$
& $<0.53$
\\
\multicolumn{2}{l}{ ($h = 0.72 \pm 0.08$) + age}\\
& MADCAP
& $1.00^{0.04}_{0.04}$
& $1.02^{0.08}_{0.07}$
& $0.023^{0.003}_{0.003}$
& $0.13^{0.04}_{0.03}$
& $0.64^{0.11}_{0.14}$
& $0.38^{0.13}_{0.13}$
& $0.053^{0.016}_{0.016}$
& $0.66^{0.09}_{0.09}$
& $13.7^{1.3}_{1.3}$
& $<0.49$
\\
& FASTER
& $0.99^{0.04}_{0.05}$
& $1.06^{0.09}_{0.07}$
& $0.023^{0.003}_{0.003}$
& $0.14^{0.05}_{0.04}$
& $0.62^{0.12}_{0.17}$
& $0.39^{0.14}_{0.14}$
& $0.055^{0.016}_{0.016}$
& $0.67^{0.09}_{0.09}$
& $13.4^{1.3}_{1.3}$
& $<0.49$
\\
\multicolumn{2}{l}{Flat+ Weak $h$ + LSS + SN}\\
& MADCAP
& (1.00)
& $1.01^{0.06}_{0.05}$
& $0.022^{0.002}_{0.002}$
& $0.12^{0.02}_{0.01}$
& $0.69^{0.04}_{0.06}$
& $0.31^{0.05}_{0.05}$
& $0.047^{0.005}_{0.005}$
& $0.69^{0.04}_{0.04}$
& $13.6^{0.4}_{0.4}$
& $<0.30$
\\
& FASTER
& (1.00)
& $1.04^{0.06}_{0.06}$
& $0.024^{0.003}_{0.002}$
& $0.12^{0.01}_{0.01}$
& $0.70^{0.05}_{0.05}$
& $0.30^{0.05}_{0.05}$
& $0.048^{0.005}_{0.005}$
& $0.70^{0.05}_{0.05}$
& $13.6^{0.4}_{0.4}$
& $<0.31$
\\
\enddata
\tablecomments{\small
Cosmological parameter estimates for the FASTER and MADCAP results,
derived using a series of more restrictive applied priors.
The results show remarkable consistency between the two analysis 
paths, for all priors.  The least stable parameter is $n_s$, with
fairly consistent 1/2 $\sigma$ variations between the 
two results.
}
\end{deluxetable}

\section{Conclusions}

In this paper we have presented an analysis of 50\% more data 
from the 1998 Antarctic flight of \boom than previously treated.
Our analysis is the most thorough to date, using two very different 
power spectrum estimation pipelines to derive the angular power 
spectrum of the cosmic microwave background radiation.  The two 
methods show good agreement and,
with the greater amount of data used, an increase in the precision of
measured power spectrum.  In particular, features in the power 
spectrum beyond the first peak (at $\ell \sim 200$) are 
detected with greater confidence.  Given that such features are a natural
consequence of standard cold dark matter dominated cosmological models
with adiabatic initial density perturbations, their presence
gives us greater confidence in the validity of that model set.

Within the context of these models we have estimated the 
the values of cosmological parameters using the results from both 
of our analysis methods.  The resulting parameter values are
insensitive to the small differences between the two results.
At the increased precision with which we determine the
cosmological parameters, we find that our results remain completely
consistent with a flat $\Lambda$-CDM cosmology.

\acknowledgments

The \boom project has been supported by NASA, NSF-OPP, and NERSC in the
U.S., by PNRA, Universit\'a ``La Sapienza'', and ASI in Italy, by PPARC
in the UK, and by the CIAR and NSERC in Canada.  

\bibliographystyle{apj}
\bibliography{cmbr_jr}

\end{document}